\documentclass[aps,superscriptaddress,showpacs,nofootinbib,eqsecnum,prd,notitlepage]{revtex4} 

\usepackage{epsf}  
\usepackage{eucal} 
\usepackage{amssymb}
\usepackage{amsmath} 
\usepackage{graphicx} 
\usepackage{bm}
\usepackage{dcolumn} 
\usepackage{epsfig} 
\usepackage{multirow}
\usepackage{hyperref}
\usepackage{color}
\usepackage{float}

\begin{document}

\title{Chiral phase transition in a planar four-Fermi model in a tilted magnetic field}

\author{Rudnei O. Ramos} \email{rudnei@uerj.br} \affiliation{Departamento de
  F\'{\i}sica Te\'orica, Universidade do Estado do Rio de Janeiro, 20550-013
  Rio de Janeiro, RJ, Brazil}

\author{Pedro H. A. Manso} \email{phmansophy@gmail.com}
\affiliation{Departamento de F\'{\i}sica Te\'orica, Universidade do
  Estado do Rio de Janeiro, 20550-013 Rio de Janeiro, RJ, Brazil}

\begin{abstract}

We study a planar four-Fermi Gross-Neveu model in the presence of a
tilted magnetic field, with components parallel and perpendicular to
the system's plane. We determine how this combination of magnetic
field components, when applied simultaneously, affects the phase
diagram of the model.  It is shown that each component of the magnetic
field causes a competing effect on the chiral symmetry  in these
fermionic systems.  While the perpendicular component of the magnetic
field tends to make the chiral symmetry breaking stronger,
the effect of the parallel component of the field  in these planar
systems is to weaken the chiral symmetry through the enhancement of 
the Zeeman energy term. We show that this competing
effect, when combined also with temperature and chemical potential,
can lead  to a rich phase diagram, with the emergence of multiple
critical points and reentrant phase transitions.  We also study how
the presence of these multiple critical points and reentrant phases
can manifest in the quantum Hall effect. Our results provide a
possible way  to probe experimentally chiral symmetry breaking and the
corresponding emergence of a gap (i.e., the presence of a nonvanishing
chiral  vacuum expectation value) in planar condensed matter systems
of current interest.

\end{abstract}


\maketitle

\section{Introduction}
\label{sec1}

The interaction of fermions with an external magnetic field is
expected to be associated with phenomena like metal-to-insulating
phase transitions in  semiconductors~\cite{mit}, the quantum Hall
effect~\cite{hall} and transport properties in
superconductors~\cite{supercond},  just to mention a few examples in
the context of condensed matter physics. Among these systems, planar
ones are of particular interest and they include, for example,
high-temperature superconductors, organic thin films and most
recently, the physics of graphene~\cite{Graphene}.  Most of these
systems have an excitation spectrum that can be well described by
relativistic Dirac-like fermions.  A typical example is the physics of
graphene, in which electron transport is essentially governed by the
Dirac's relativistic equation.  It is then expected that these types of
planar systems can be appropriately  described by quantum field theory
models in two spatial dimensions.  In this context, the Gross-Neveu (GN)
type of models~\cite{GN} in $2+1$ dimensions are very popular  not
only because of their simplicity, but also for their ability to
capture many of the relevant physics exhibited by planar fermionic
systems in general. These models can have either a discrete, $\psi \to
\gamma_5 \psi$, or a continuous, $\psi \to \exp(i \alpha \gamma_5)
\psi$, chiral symmetry.  They consist of $N$ massless four-component
fermions self-interacting  through a local four-Fermi interaction. 

One of the most basic motivations for studying the properties of
fermionic planar systems is to determine whether we can have a 
metal-to-insulating
type of transition  under the variation of external parameters, like
temperature, chemical potential and external fields, e.g., an external
magnetic field.  The metal-insulator transition is directly connected
to the presence of a nonvanishing gap in the excitation spectrum for
the fermions and how this gap eventually vanishes in the presence of
external conditions.  In the quantum field theory context,  a mass
term for the fermions can be generated dynamically in the absence of
external conditions, like in the original GN model,   and it is
associated with the breaking of the chiral symmetry in the model.

\begin{figure}[thb]
\centerline{\psfig{file=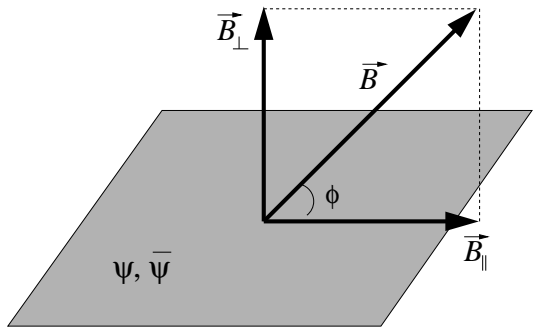,width=7cm}}
\caption{A schematic representation of the system under study,
  consisting of fermions in a plane and in the presence of a magnetic
  field $\vec B$ that makes an angle $\phi$ with respect to the
  plane.}
  \label{BBfig1}
\end{figure}

In this work we make use of the original GN model with discrete chiral
symmetry and study the chiral phase transition when the system is in
the presence of an external  tilted magnetic field, i.e., a magnetic
field that includes both parallel and perpendicular components that
are being applied simultaneously to the system (see
{}Fig. \ref{BBfig1}).  This model has been previously extensively
analyzed when in the presence of a magnetic field applied
perpendicular to the plane (see, for example,
Refs.~\cite{Klimenko,Gusynin} for some of the earlier studies) and,
more recently, in the presence of an in-plane magnetic
field~\cite{PRBinplane}.  But, surprisingly, a study of both types of
fields, parallel and perpendicular to the system's plane, applied
simultaneously  is still lacking.  Perhaps the reason for this
negligence in the literature was because not much was expected for the
effects on the system symmetry due to a parallel component of the
field.  However, there are good reasons for making this study. It is
well known that a perpendicular magnetic field $B_\perp$ leads to an
{\it enhancement} of the chiral-symmetry-broken region (thus, it tends
to increase a gap in the excitation spectrum). This is
the magnetic catalysis phenomenon that can occur for all values of the
four-Fermi interaction  strength~\cite{Klimenko,Gusynin}. However, an
in-plane magnetic field $B_\parallel$ has been shown to produce an
opposite effect~\cite{PRBinplane}, leading to a {\it reduction} of the
chiral symmetry broken region. In other words, a parallel component of
the magnetic field tends to act on restoring the chiral symmetry for
large enough values of $B_\parallel$. This is due to the fact that
the in-plane magnetic field contributes to enhance the
Zeeman energy term (which depends on the total magnetic field applied to the system), 
which in turn leads to larger effective polarization of the
system. The polarized  planar fermionic system tends to exhibit less
chiral symmetry.

Since the application of perpendicular and parallel (in-plane)
magnetic fields leads to opposite effects on the chiral symmetry, it
is reasonable to expect the combination of both components, when
applied  simultaneously to the system, to possibly lead to new
patterns for the chiral  phase transition. We will show in this work that
this indeed can happen under an appropriate choice of parameters. 
In this paper we show that the
application of both $B_\perp$ and $B_\parallel$, with  the combined
effects of temperature $T$ and  chemical potential $\mu$, can lead to
a phase diagram that can display a  rich structure, with the
possibility of exhibiting multiple critical points and reentrant
phases.  We show that this is a consequence of the competing effects
on the system's  chiral symmetry coming from each component of the
applied oblique external magnetic  field.

This paper is organized as follows: In Sec.~\ref{sec2} we briefly
revise the four-Fermi GN model used in this work.  In Sec.~\ref{sec3}
the thermodynamic effective potential for the fermionic  scalar
condensate  $\langle {\bar \psi} \psi \rangle$ is  derived in the
mean field theory approximation, and results for the phase diagram for
the model are displayed.  In Sec.~\ref{sec4} we study in particular
the quantum phase transitions in the model and derive the Hall
conductivity. We also  discuss the impact that the presence of the 
possible multiple critical points can have on the Hall conductivity, 
pointing out a possible way for characterizing  experimentally the presence of
quantum critical points.  {}Finally, in Sec.~\ref{sec5}, we give our
conclusions and discuss  possible applications and extensions of this
work.


\section{The Four-Fermi Model in the Presence of an External Magnetic Field}
\label{sec2}

We work with a four-Fermi model, in $2+1$ dimensions, of the GN type
and with $N$ flavors, described by the Lagrangian density~\cite{GN}

\begin{equation}
{\cal L} = \sum_{s=\uparrow,\downarrow} \sum_{j=1}^N \left[
  \bar{\psi}_j^s (i \not\!\partial)  \psi_j^s + \frac {\lambda}{2 N}
  (\bar{\psi}_j^s\psi_j^s)^2 \right]\;,
\label{Lag4F}
\end{equation}
where $\lambda$ is the  coupling and $\not\!\!\partial \equiv
\gamma^\nu \partial_\nu$, $\nu=0,1,2$, with the gamma matrices being
$4\times 4$ matrices, where here we follow the  representation given,
e.g., in Ref.~\cite{Park} for fermions in 2+1dimensions.  The model of
Eq.~(\ref{Lag4F}) possesses a discrete chiral symmetry, $\psi \to \gamma_5
\psi$, $\bar{\psi} \to - \bar{\psi} \gamma_5$, with the $\gamma_5$
matrix defined as in Ref.~\cite{Park}. This discrete chiral symmetry
is the one considered in this paper, along with its breaking and
restoration conditions.  The chiral symmetry is broken when a gap (a
mass term for the fermions) is generated.  Since the gap corresponds
to a nonvanishing  vacuum expectation value for the chiral operator,
$\langle \bar{\psi} \psi \rangle$, it is then convenient to rewrite
Eq.~(\ref{Lag4F}) in terms of an auxiliary scalar field $\sigma$,  

\begin{eqnarray}
{\cal L}[\bar{\psi},\psi, \sigma] =  \sum_{s=\uparrow,\downarrow}
\sum_{j=1}^N  \bar{\psi}_j^s (i \not\!\partial -\sigma)  \psi_j^s -
\frac{N}{2 \lambda}  \sigma^2\;,
\label{LagTLM}
\end{eqnarray}
where $\sigma$ and the chiral operator are related, from the
saddle-point solution for $\sigma$, by $\sigma= -(\lambda/N)
\bar{\psi}\psi$, with implicit sums over the spin and flavor.  In
terms of the Lagrangian density (\ref{LagTLM}), the discrete chiral
symmetry is now expressed as  $\psi \to \gamma_5 \psi$, $\bar{\psi}
\to - \bar{\psi} \gamma_5$ and $\sigma \to - \sigma$.

With the Lagrangian density of Eq.~(\ref{LagTLM}) expressed in terms of
the auxiliary scalar field, it becomes relatively simpler to study the
phase diagram of the model, which can be made through the study of the
behavior of the vacuum expectation value of $\sigma$ as a function of
the temperature, the chemical potential and the  external magnetic
field. This study can be made by evaluating the effective potential,
or Landau's free energy density, for a constant (background)
configuration for the auxiliary scalar field, $V_{\rm eff}
(\sigma_c)$.  The effective potential can be  obtained by integrating
out the fermion fields and fluctuations around the  scalar background
field~\cite{Park}. {}From the effective potential, all the
thermodynamics for the model can be derived.  In particular, a chiral
phase transition is signaled by a nonvanishing vacuum expectation
value for $\sigma$, ${\bar \sigma}_c \equiv \langle \sigma \rangle$,
which is a (global) minimum of $V_{\rm eff} (\sigma_c)$. Both
temperature and chemical potential are introduced through the grand
canonical partition function definition,

\begin{equation} 
Z(\beta,\mu) = {\rm Tr} \exp\left[ - \beta \left(  H - \mu Q \right)
  \right] \;, 
\label{Zbetamu} 
\end{equation} 
where $\beta$ is the inverse of the temperature\footnote{Throughout this
  work, we assume  the natural units $k_B=\hbar=c=1$.}, $\mu$ is the
chemical potential, $H$ is the Hamiltonian corresponding to
Eq.~(\ref{LagTLM}) and  $Q=\int d^2 x \bar{\psi} \gamma_0 \psi$ is the
conserved charge.  {}Furthermore, we can also study the effect of an
external magnetic field applied to the system by coupling it to the
fermions in Eq.~(\ref{LagTLM}) in the usual way. {}For an arbitrary
external magnetic field that can have components both parallel and
perpendicular to the system's plane,  the perpendicular component of the field
couples to the  fermions orbital motion, generating the Landau levels
for fermions in a magnetic field~\cite{Klimenko,Gusynin}, while  the
total component of the field contributes to an intrinsic Zeeman 
effect~\cite{Kittel}.
The effective potential, or free energy density, is defined as usual
from the grand canonical partition function Eq.~(\ref{Zbetamu}) by

\begin{eqnarray}
V_{\rm eff}=-\frac{1}{{\beta \cal V}} \ln Z \;,
\label{veff0}
\end{eqnarray}
where  ${\cal V}$ is the volume (or, more specifically, the area  in
the present problem of a planar system).

Transforming Eq.~(\ref{Zbetamu}) in the form of a path integral, in
the imaginary-time (Euclidean) formalism of finite temperature field
theory~\cite{kapusta},  one obtains

\begin{equation}
Z=\int D \sigma \prod_{s} D \psi^{\dagger}D \psi \, \exp  \left\{ -
S_E[\bar{\psi},\psi,\sigma] \right\} ,
\label{partfun}
\end{equation}
where the functional integration is performed over the fermion fields
satisfying the antiperiodic boundary condition in Euclidean time:
$\psi_j(x,\tau) = - \psi_j (x,\tau + \beta)$, while over the $\sigma$
it satisfies periodic boundary conditions. The Euclidean action
$S_E[\bar{\psi},\psi]$ in Eq.~(\ref{partfun}),  from the Lagrangian
density of Eq.~(\ref{LagTLM}) in the presence of an external magnetic
field with components $B_\perp$ and $B_\parallel$, perpendicular and
parallel to the plane, respectively, is given by\footnote{Note that in
  condensed matter systems there should also be a Fermi velocity $v_F$
  factor multiplying the space components of the gamma matrices and
  the coupling constant $\lambda$ in this expression~\cite{PRBinplane}.
  (For example, in graphene $v_F \approx c/300$). $v_F$ is omitted in
  all our expressions,  but we will reestablished it when making
  estimates in the context of condensed matter systems later on.}

\begin{eqnarray}
S_E[\bar{\psi},\psi,\sigma] = && \int_0^{\beta} d\tau \int d^2x~
\left\{  \sum_{s=\uparrow,\downarrow} \sum_{j=1}^N  \bar{\psi}_j^s
\left[ \gamma_0 \partial_\tau  + i  \gamma_1  \left(\partial_x +  i
  e\, A_{x}  \right)  +  i \hbar \gamma_2  \left(\partial_y +  i e\,
  A_{y}  \right) \right. \right. \nonumber \\ && \left. \left. +
  \sigma +  \gamma_0 \mu + \frac{\sigma_s}{2} \gamma_0 \, g \, \mu_B
  B \right] \psi_j^s + \frac{N}{2 \lambda}  \sigma^2
\right\}\;,
\label{action}
\end{eqnarray}
where $A_{x}$ and $A_{y}$ are the external electromagnetic  vector
potential  components, $\sigma_s \gamma_0 \, g \, \mu_B
B/2$, with $B \equiv |\vec{B}| = \sqrt{B_\perp^2 + B_\parallel^2}$, 
is the corresponding Zeeman energy term; 
$\sigma_\uparrow =1$, $\sigma_\downarrow =-1$;  $g$ is the
spectroscopic Lande factor and $\mu_B$ is the Bohr magneton,
$\mu_B = e^2/(2 m_e)$.  By
choosing a gauge where the three-dimensional vector potential is
given, for example, by $\vec{A} = (0,B_\perp x,B_\parallel y)$, we see
from Eq.~(\ref{action}) that $B_\perp$ couples only to the orbital
motion of the fermions and will result in
the Landau levels for the system in this magnetic field. The Zeeman
energy term in Eq.~(\ref{action}) involves the total
magnitude of the magnetic field and originates from the standard form for
the Zeeman term as employed in quantum many-body theory, 
for instance~\cite{Kittel}. 
Note also that from the form of the Zeeman
energy term in Eq.~(\ref{action}), we see that it can be added to the
chemical potential.  Thus, in the action $S_E$, we can define an
effective chemical potential term of the form,

\begin{eqnarray}
\sum_{s=\uparrow,\downarrow} \sum_{j=1}^N\mu_s \bar{\psi}_j^s \gamma_0
\psi_j^s &=& \sum_{s=\uparrow,\downarrow}  \sum_{j=1}^N \left(\mu +
\frac{\sigma_s}{2} \, g \, \mu_B B\right) \bar{\psi}_j^s
\gamma_0 \psi_j^s \nonumber \\ &=& \mu_{\uparrow}
      {\psi^\uparrow}^{\dagger}  \psi^{\uparrow} + \mu_{\downarrow}
      {\psi^\downarrow}^{\dagger}  \psi^{\downarrow}\;,
\label{asym}
\end{eqnarray}
where $\mu_\uparrow= \mu + \delta \mu$ and $\mu_\downarrow= \mu -
\delta \mu$, with $\delta \mu = g \, \mu_B B/2$. The effect
of the total amplitude of the magnetic field is then felt through
the asymmetry term $\delta \mu$ in the chemical potential for spin up
and spin down, resulting in a polarization of the system. 
Note that in the language of condensed matter systems,
the  magnetic field produces the analogue of an asymmetrical
doping, or an imbalance between the chemical potentials of the  fermions
(electrons) with the two possible spin orientations. This asymmetry can 
be produced by a doping process that alters the
densities of spin up and spin down. Physically, the chemical potential
$\mu$  can be interpreted as to account for the extra density of
electrons that  is supplied to the system by the dopants, while
$\delta \mu$ 
measures the amount of asymmetry introduced.
We should also note that even in the absence of the parallel component
of the magnetic field, $B_\parallel=0$, there is still a Zeeman
energy term, which is then proportional only to $B_\perp$. However,
in most practical applications this Zeeman energy contribution, when
compared with the lowest Landau energy level, $\sqrt{2 |e B_\perp|}$,
is very small
and can safely be neglected (see the discussion later on in the 
Conclusions section).
The Zeeman energy term is then, e.g. in practical physical systems
found in condensed matter systems, only appreciable when
$B_\parallel \gg B_\perp$. Expressing the Zeeman energy term 
in terms of the tilt angle $\phi$ of the
magnetic field (see {}Fig.~\ref{BBfig1}), $\delta \mu = 
g \, \mu_B B/2 = g \, \mu_B B_\parallel \sqrt{1+ \tan^2 \phi}/2$, we
see that the 
importance of the Zeeman energy term will in general be 
when $\phi \ll 1$, or $\delta \mu  \approx  g \, \mu_B B_\parallel/2$,
corresponding to a highly tilted magnetic field\footnote{Alternatively,
instead of a highly tilted magnetic field, we could also think
in a physical situation where $B_\perp$ and $B_\parallel$ are
generated independently and such that $B_\parallel \gg B_\perp$.}.
Under these circumstances, the polarization 
produced by the Zeeman energy term then becomes proportional to
$B_\parallel$. 

In the analysis we will perform in the following sections
on the effect
of the magnetic field on the phase structure of the GN model,
we will work with the magnetic fields expressed in terms
of their contributions through the Landau energy levels 
(given in terms of $B_\perp$) and in terms of the asymmetry
$\delta \mu=g \, \mu_B B_\parallel \sqrt{1+ \tan^2 \phi}/2$.
We also keep
in mind that for typical physical problems of relevance
in condensed matter systems, in order for the asymmetry (Zeeman term)
to be strong enough such that it cannot
be neglected (when compared with the magnitude of the Landau energy terms), 
it is required in general that  $B_\parallel \gg B_\perp$,
or, equivalently, $\phi \ll 1$, which corresponds to a highly tilted magnetic field
(this is also discussed in more detail in the Conclusions section).


\section{Effective potential and phase structure}
\label{sec3}

All the thermodynamics, magnetic properties and phase structure of the
model Eq.~(\ref{LagTLM}), as a function of temperature, chemical potential
and magnetic field,  can be determined from the effective potential
Eq.~(\ref{veff0}).  It is useful to start by briefly reviewing the
derivation of the effective potential and its main results for the GN
model, including at first only temperature and chemical potential and
then later also including the external magnetic field.


\subsection{The effective potential and phase structure at finite $T$ and $\mu$}

{}From the grand canonical partition function Eq.~(\ref{partfun}),
with the action (\ref{action}) (considering it initially in the absence of
an external magnetic field),
for a constant background scalar field $\langle
\sigma\rangle=\sigma_c$, in the mean field approximation, or
equivalently, from the leading term in a $1/N$ expansion, or the
large-$N$ approximation~\cite{coleman,largeNreview}, we obtain that
the effective potential $V_{\rm eff}$ for $\sigma_c$, at finite $T$
and $\mu$, is given by 

\begin{widetext}
\begin{eqnarray}
V_{\rm eff} (\sigma_c,T, \mu)=  \frac{N}{2\lambda}  \sigma_c^2   - 2 N
T  \sum_{n=-\infty}^{+\infty} \int \frac{d^2 p}{(2 \pi)^2} \ln
\left[\left(\omega_n -i  \mu \right)^2 +{\bf p}^2 +\sigma_c^2 \right]
\;,
\label{Veff1}
\end{eqnarray}
\end{widetext}
where $\omega_n=(2n+1)\pi  T$ are the Matsubara frequencies for
fermions.  Performing the sum  over the Matsubara  frequencies in
Eq.~(\ref{Veff1}), we find 

\begin{eqnarray}
V_{\rm eff}(\sigma_c,T, \mu) &=& \frac{N}{2 \lambda}  \sigma_c^2  - 2
N  \int\frac{d^2p}{(2\pi)^2 } E_p   \\ &-& 2 N T
\int\frac{d^2p}{(2\pi)^2 }\left[  \ln\left(1+e^{-\beta E^+}\right)  +
  \ln \left(1+e^{-\beta E^-}\right)  \right]\;,
\label{poteff}
\end{eqnarray}
where $E^{\pm} = E_p \pm \mu$ and $E_p=\sqrt{{\bf p}^2+\sigma_c^2}$. 

At zero temperature and chemical potential the effective potential
becomes

\begin{eqnarray}
V_{\rm eff}(\sigma_c)=\frac{N}{ 2 \lambda} \sigma_c^2 - 2 N \int
\frac{d^2p}{(2\pi)^2}~  \sqrt{{\bf p}^2 +\sigma_c^2}\;.
\label{poteffTmu0}
\end{eqnarray}

The momentum integral in Eq.~(\ref{poteffTmu0}) is ultraviolet
divergent.  By using a momentum cutoff $\Lambda$ to regulate the
divergent term in $V_{\rm eff}(\sigma_c)$, we can define a
renormalization condition for  the coupling as

\begin{equation}
\frac{1}{\lambda_R(m)} =\frac{1}{N}  \frac{d^2 V_{\rm
    eff}(\sigma_c)}{d \sigma_c^2}\Bigr|_{\sigma_c=m}\;,
\label{renor}
\end{equation}
where $m$ is a regularization scale. Equation (\ref{renor}) gives
(neglecting terms of ${\cal O}(1/\Lambda)$),

\begin{equation}
\frac{1}{\lambda_R(m)} = \frac{1}{\lambda} - \frac{\Lambda}{\pi} +
\frac{2m}{\pi}\;.
\end{equation}
Next, by defining a renormalization invariant coupling 
as~\cite{Vshivtsev:1996ri} 

\begin{equation}
\frac{1}{\lambda_R} =  \frac{1}{\lambda_R(m)} - \frac{2m}{\pi}\;,
\end{equation}
we obtain that in terms of $\lambda_R$ the renormalized effective
potential reads (after subtracting an irrelevant field-independent
divergent vacuum term)

\begin{eqnarray}
V_{\rm eff,R}(\sigma_c) = \frac{N}{ 2  \lambda_R} \sigma_c^2 +
\frac{N}{3\pi} |\sigma_c|^3\;.
\label{poteffR}
\end{eqnarray}
The minimum $\bar{\sigma}_c$ of $V_{eff,R}(\sigma_c)$ is given by

\begin{eqnarray}
\bar{\sigma}_c=\begin{cases} \sigma_0 \ , & \lambda_R<0,\\ 0 \ , &
\lambda_R>0\,,
\end{cases}
\end{eqnarray}
where $\sigma_0=\pi/|\lambda_R|$.  Thus, dynamical chiral symmetry
breaking, in the absence of external fields, only occurs for a
negative coupling constant.

Using (\ref{poteffR}) and performing the (finite) momentum integrals
for the temperature- and chemical-potential-dependent terms in
Eq.~(\ref{poteff}), the renormalized effective potential at finite $T$
and $\mu$ becomes

\begin{eqnarray}
V_{\rm eff}(\sigma_c,T, \mu) &=& V_{\rm eff,R}(\sigma_c)  \nonumber
\\ &+&  \frac{N }{\pi} T^2 |\sigma_c|  \left\{ {\rm
  Li}_2[-e^{-\beta(\sigma_c-\mu)}]  +  {\rm
  Li}_2[-e^{-\beta(\sigma_c+\mu)}]  \right\} \nonumber \\ &+&
\frac{N}{\pi}T^3  \left\{ {\rm Li}_3[-e^{-\beta(\sigma_c-\mu)}] +
     {\rm Li}_3[-e^{-\beta(\sigma_c+\mu)}] \right\} \;,     
\label{veffTmu}
\end{eqnarray}
where  ${\rm Li}_\nu(z)$ is the polylogarithm function, and it is
defined (for $\nu >0$) as \cite{Abramowitz}

\[
{\rm Li}_\nu(z)= \sum_{k=1}^{\infty} \frac{z^k}{k^\nu}\;.
\]

{}From Eq.~(\ref{veffTmu}), we can derive  the gap equation, which is
defined from

\begin{equation}
\frac{\partial}{\partial \sigma_c}V_{\rm eff}(\sigma_c,T, \mu)
\Bigr|_{\sigma_c = \bar{\sigma}_c(T, \mu)} =0\;,
\label{gap}
\end{equation}
which gives

\begin{eqnarray}
\bar{\sigma}_c &=& \bar{\sigma}_c(T=0, \mu=0) -
\frac{1}{\beta}\left\{ \ln \left[ 1+ e^{-\beta(\bar{\sigma}_c+
    \mu)}\right]  + \ln \left[ 1+ e^{-\beta(\bar{\sigma}_c-
    \mu)}\right]  \right\}\;.
\label{Deltacgap}
\end{eqnarray}

The behavior of $\bar{\sigma}_c$ is well known~\cite{Park,rose}.  At
$T=0$, there is a critical chemical potential $\mu_c$ at which the
chiral symmetry is restored (we are considering the case here where
chiral symmetry breaking happens, i.e., for $\lambda_R<0$).   This can
be shown to happen through a first-order phase transition, defined by
the relation $V_{\rm eff,R}(\sigma_c=0,\mu_c)=V_{\rm
  eff,R}(\sigma_c=\sigma_0,\mu_c)$.  This yields the critical chemical
potential $\mu_c=\sigma_0$.  Thus, at zero temperature, we have the
ground state of the system given by
 
\begin{eqnarray}
\label{gs1}
\bar{\sigma}_c= \left\{
\begin{array}{cc}
\sigma_0, &{\rm for}~~\mu<\mu_c\;,\\ 0,&{\rm for}~~\mu \geq \mu_c\;.
\end{array}
\right.
\end{eqnarray}
At finite temperature and chemical potential, there is a critical
curve  $\bar{\sigma}_c(T,\mu)=0$, which is obtained from
Eq. (\ref{Deltacgap}). This gives a line of second-order phase
transition in the $(\mu,T)$ plane, starting at the critical point
$(\mu=0,T=T_c)$, where $ T_c = \sigma_0/(2 \ln 2)$,  and ending in a
first-order critical point at $(\mu=\mu_c,T=0)$. Note that when going
beyond the  mean field approximation, this structure can slightly
change~\cite{gn3d}, leading to a first-order critical line merging
with the second-order critical line at a tricritical point. In this
work, we will restrict our analysis at the mean field approximation
only.


\subsection{The effective potential for the GN model at finite $T,\,\mu$ and
in a tilted magnetic field}

Let us now investigate the effective potential for the system   in the
presence of a tilted magnetic field, with components
$(B_\parallel,B_\perp)$. The effective potential can once again be
derived  from Eq.~(\ref{partfun}) and it can also follow directly from
Eqs.~(\ref{Veff1}) or (\ref{poteff}). In the case where the magnetic field
has only an in-plane component, $B_\parallel\neq 0$ and $B_\perp=0$,
we have a shift of the chemical potential and we need to distinguish
in the effective potential, not only the contributions from particles
and antiparticles, but also the contributions from particles with spin up
and spin down,  with chemical potentials $\mu_\uparrow= \mu + \delta
\mu$ and  $\mu_\downarrow= \mu - \delta \mu$, respectively, and 
with $B_\perp=0$,
$\delta \mu = g \, \mu_B B_\parallel/2$.  Throughout this work we will
be considering $\mu$ and $\delta \mu$ as positive
quantities. Equation (\ref{poteff}) for
the renormalized effective potential, in this case, becomes

\begin{eqnarray}
V_{\rm eff}(\sigma_c,T, \mu_\uparrow,\mu_\downarrow) &=& \frac{N}{ 2
  \lambda_R} \sigma_c^2 + \frac{N}{3\pi} |\sigma_c|^3  \nonumber
\\ &-& N T \int\frac{d^2p}{(2\pi)^2 }\left[  \ln\left(1+e^{-\beta
    E_\uparrow^+}\right)  + \ln \left(1+e^{-\beta E_\uparrow^-}\right)
  \right] \nonumber \\ &-& N T \int\frac{d^2p}{(2\pi)^2 }\left[
  \ln\left(1+e^{-\beta E_\downarrow^+}\right)  + \ln \left(1+e^{-\beta
    E_\downarrow^-}\right)  \right]\;,
\label{veffTmudeltamu}
\end{eqnarray}
where $E_{\uparrow,\downarrow}^{\pm} = E_p \pm
|\mu_{\uparrow,\downarrow}|$.  A detailed analysis of the phase
structure in this case was performed in Ref.~\cite{PRBinplane}. The
main effect of the in-plane magnetic field is to further contribute
towards chiral symmetry restoration.  In this case, it can be shown
that the chiral symmetry will remain always  restored (at any $T$ and
$\mu$) for a critical asymmetry  $\delta \mu_c \geq \mu_c =
\sigma_0$.

{}Finally, we can also include in the above equation
(\ref{veffTmudeltamu}) a perpendicular component for the magnetic
field, $B_\perp \neq 0$, by properly accounting for the Landau energy
levels and degeneracies. {}For instance, in Eq.~
(\ref{veffTmudeltamu}), we can  take~\cite{Klimenko,Gusynin} $E_p =
\sqrt{{\bf p}^2 + \sigma_c^2} \to \sqrt{2k | e B_\perp| +
  \sigma_c^2}$, $k=0,1,2,\ldots$, the momentum integrals are replaced
by a sum over the Landau levels, with a density of states
$|eB_\perp|/(2 \pi)$ and by also accounting for the  degeneracy of the
$k \geq 1$ Landau levels, we obtain

\begin{eqnarray}
\frac{1}{N} V_{\rm eff,R}(\sigma_c,T,\mu,B_{\perp},\delta \mu) &=&
\frac{\sigma_c^2}{2 \lambda_R} -  \frac{\sqrt{2} |e
  B_\perp|^{3/2}}{\pi} \zeta\left( -\frac{1}{2}, \frac{\sigma_c^2}{2
  |e B_\perp|} +1 \right) - \frac{|\sigma_{c}||eB_{\perp}|}{2 \pi}
\nonumber \\ &-& \frac{|eB_{\perp}|}{4\pi\beta} \left\lbrace  \ln
\left( 1+e^{-\beta(|\sigma_{c}|-\mu_{\uparrow})}\right)  +  \ln \left(
1+e^{-\beta(|\sigma_{c}|+\mu_{\uparrow})}\right) \right.  \nonumber
\\ &+& \left.  2\sum^{\infty}_{k=1}\ln \left( 1+e^{-\beta \left(
  \sqrt{\sigma_{c}^2+2k|eB_{\perp}|}-\mu_{\uparrow}\right)} \right)
\right.   \nonumber \\ &+& \left.  2\sum^{\infty}_{k=1}\ln \left(
1+e^{-\beta \left(  \sqrt{\sigma_{c}^2+2k
    |eB_{\perp}|}+\mu_{\uparrow}\right)} \right) \right.  \nonumber
\\ &+& \left.  (\mu_{\uparrow} \rightarrow
|\mu_{\downarrow}|)\frac{}{}\right\rbrace,
\label{Veff}
\end{eqnarray} 
where $\zeta(s,a)$ is the Hurwitz zeta function~\cite{zeta},

\begin{equation}
\zeta(s,a) = \sum_{k=0}^{\infty} \frac{1}{(k+a)^s}\;.
\label{zetafunc}
\end{equation}

Equation (\ref{Veff}) can be seen as a  generalization of the
mean field theory (or large-$N$ approximation) result derived, e.g.,
in  Refs.~\cite{Gusynin,Vshivtsev:1996ri}, for the case considered here, 
where we have a parallel component of the magnetic field (or equivalently, 
an asymmetry between the spin-up and spin-down components
of the fermion field).

An important feature concerning the (2+1)-dimensional GN model in a
magnetic field, when $B_\perp \neq 0$ and taking $B_\parallel=0$, is
that it can be shown that a  nonvanishing global minimum can always
develop for the effective potential, even when the coupling in
Eq.~(\ref{Veff}) is positive.  This is the phenomenon of {\it magnetic
  catalysis},  found and studied at length, e.g., in
Refs.~\cite{Klimenko,Gusynin}.  The behavior for the global minimum
$\bar{\sigma}_c$ for small values of $B_\perp$, at
$B_\parallel=T=\mu=0$, has been determined to be given
by~\cite{Klimenko}

\begin{eqnarray}
\bar{\sigma}_c=\begin{cases}
\frac{1}{2\pi}|\lambda_R||eB_{\perp}|+\cdots \ , &
\lambda_R>0,\\ \sigma_0\left[1+\frac{|eB_{\perp}|^{2}}{12
    \sigma_0^{4}}+\cdots \right] \ , & \lambda_R<0.
\end{cases}
\end{eqnarray}
Thus, the component of the magnetic field perpendicular to the plane
always tends to break (or enhance) the chiral symmetry.  On the other
hand, as we have already commented above, a  parallel (in-plane)
magnetic field component $B_\parallel \neq 0$ has been
shown~\cite{PRBinplane}  to always tend to restore the chiral
symmetry.  Therefore, when both components of the magnetic field are
present, one can expect that there must be a competing effect  between
the  in-plane component $B_\parallel$ and the perpendicular component
$B_\perp$. While the former tends to suppress the chiral symmetry due
to the increase of the magnitude of the Zeeman energy term (the asymmetry
$\delta \mu$), the
latter tends to enhance the chiral-symmetry-broken region.  
We then expect that the interplay of the
two components of the  magnetic field,  when applied simultaneously
and independently to the system,  will possibly lead to a rich
structure for the phase diagram.  This indeed will be true, as we will
show below.  As we will show, the simultaneous and independent
application of external magnetic fields in parallel and perpendicular
to the system's plane can generate multicritical points and reentrant
phases. These are results not seen in  those cases when only one
component of the magnetic field is present.  It is quite surprising
that such study of the simultaneous effect of both components of the
magnetic field in the GN in 2+1 dimensions has not been done before,
despite the fact that this is a kind of physical situation that has
been of increasing interest in the context of experiments with
planar types of condensed matter systems in the  laboratory. We will
comment more on this in the Conclusions section.


\subsection{The phase diagram in the presence of a tilted magnetic field} 

We now turn to the study of the phase diagram for the (2+1)-dimensional
GN model, with the effective potential  given by Eq.~(\ref{Veff}).
The equation for the gap $\bar{\sigma}_c$, which is determined from
the equation defining the minimum of the effective potential,

\begin{eqnarray} 
\left.\frac{d}{d\sigma_c} V_{\rm
  eff,R}(\sigma_c,T,\mu,B_{\perp},\delta \mu)
\right|_{\sigma_c=\bar{\sigma}_c} = 0 \;,
\end{eqnarray}
and that generalizes Eq.~(\ref{Deltacgap}), becomes

\begin{eqnarray} \label{gapTmuBperpBparal}
0 &=&  \frac{1}{\lambda_{R}}\bar{\sigma}_c-\frac{\sqrt{2}}{2\pi}\,
|eB_{\perp}|^{\frac{1}{2}} \, \bar{\sigma}_c \, \zeta
\left(\frac{1}{2},\frac{\bar{\sigma}_c^{2}}{2|eB_{\perp}|}\right)+
\frac{|eB_{\perp}|}{2\pi} \nonumber \\ &&+
\frac{|eB_{\perp}|}{4\pi}\left\{
\vphantom{\frac{1}{e^{\beta\left(\sqrt{\bar{\sigma}_c^{2}+
        2k|eB_{\perp}|}+\mu_{\uparrow}\right)}+1}}
\frac{1}{e^{\beta(\bar{\sigma}_c-\mu_{\uparrow})}+1}+
\frac{1}{e^{\beta(\bar{\sigma}_c+\mu_{\uparrow})}+1}
\right. \nonumber \\ &&
\left. +2\bar{\sigma}_c\sum_{k=1}^{\infty}\frac{1}{(\bar{\sigma}_c^{2}+
  2k|eB_{\perp}|)^{\frac{1}{2}}}\left[\frac{1}{
    e^{\beta\left(\sqrt{\bar{\sigma}_c^{2}+2k|eB_{\perp}|}-\mu_{\uparrow}\right)}+1}
  \right. \right.  \nonumber \\ &&\left. \left.
  +\frac{1}{e^{\beta\left(\sqrt{\bar{\sigma}_c^{2}+2k|eB_{\perp}|}+
      \mu_{\uparrow}\right)}+1} \right]  +\left(\mu_{\uparrow}
\rightarrow |\mu_{\downarrow}|\right)  \right\} .
\end{eqnarray}

A second-order phase transition critical point is found when, by
changing any of the external parameters $T,\, \mu,\, B_\perp$ and/or
the asymmetry $\delta \mu$, $\bar{\sigma}_c$ changes continuously and
vanishes at the critical point.  {}For a first-order phase transition,
we have, instead, that $\bar{\sigma}_c$ is discontinuous at the
critical point.  We are interested in obtaining the critical points by
varying a set of parameters, while others are kept fixed. {}For
example, we can initially fix the values for the asymmetry and for
the perpendicular to the plane component of the magnetic field, while
varying the temperature and chemical potential.  {}For fixed values of
$B_\perp$ and $\delta \mu$, a second-order
transition line in the $(\mu,T)$ plane  then follows from those
critical points for which $\bar{\sigma}_c$ vanishes in a continuous
way along this critical line. In a first-order transition, the
effective potential develops different minima, $\bar{\sigma}_c^{(1)}
\neq \bar{\sigma}_c^{(2)}$, where one of them is a local minimum,
while the other is a global minimum. These minima can become degenerate
for some values of the parameters.  The first-order transition line is
then  determined by the condition of degeneracy of the minima of the
effective potential, 

\begin{equation}
V_{\rm eff,R}(\bar{\sigma}_c^{(1)},T_c,\mu_c,B_{\perp},\delta \mu) =
V_{\rm eff,R}(\bar{\sigma}_c^{(2)},T_c,\mu_c,B_{\perp},\delta \mu)\;.
\label{1stline}
\end{equation}
In general, one of the minima is the trivial solution,
$\bar{\sigma}_c=0$, though this may not be always true, depending on
the parameters chosen in Eq. (\ref{Veff}), as we will verify
explicitly below.  In all cases, the critical lines are determined
numerically, by solving the gap equation (\ref{gapTmuBperpBparal}) and
by also verifying the condition  of Eq.~(\ref{1stline}). The point where the
second-order critical line meets the first-order critical line defines
a tricritical point  in the phase diagram.

\begin{figure}[htb]
\centerline{\psfig{file=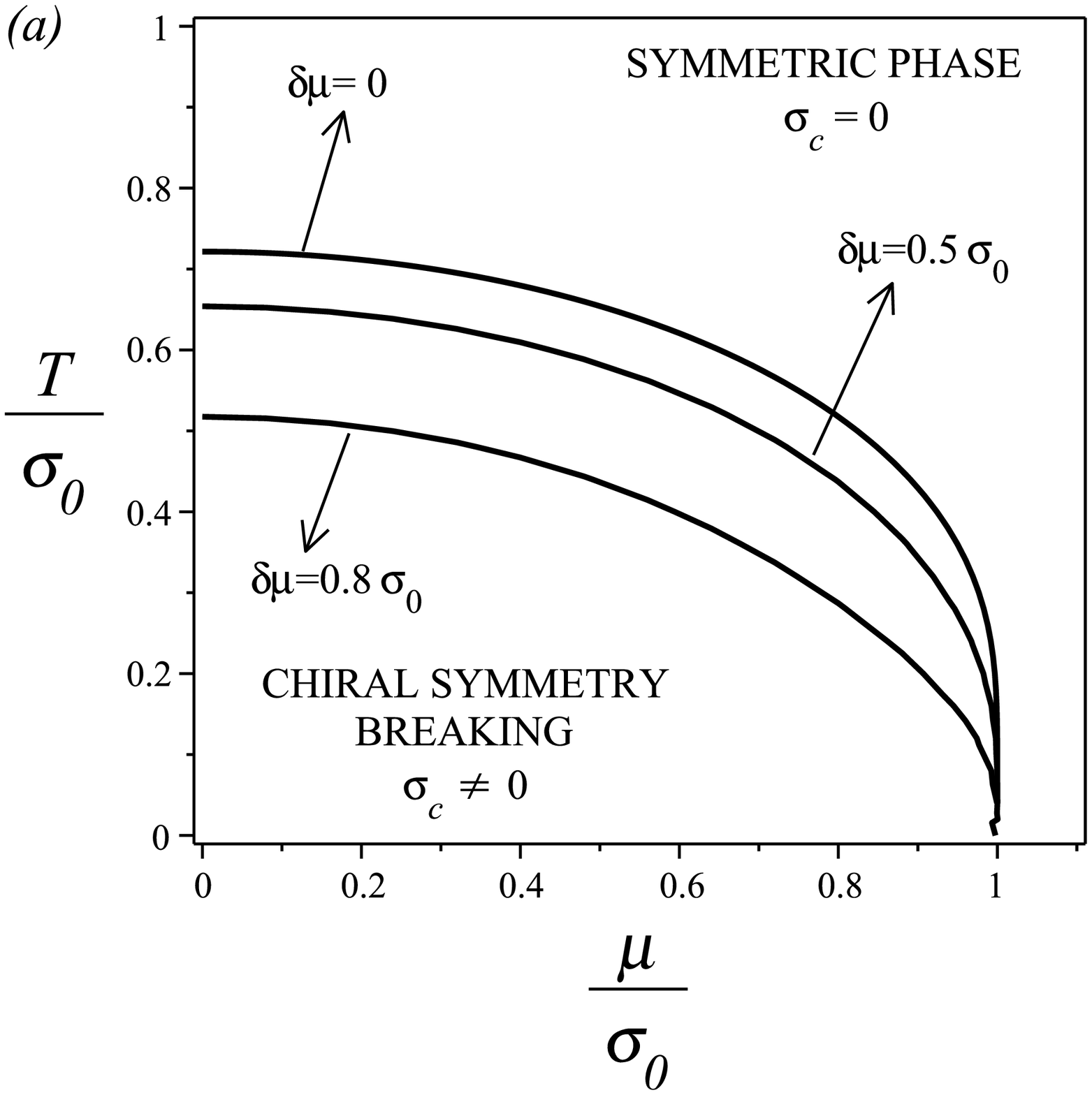,width=7.5cm}
\hspace{0.15cm} \psfig{file=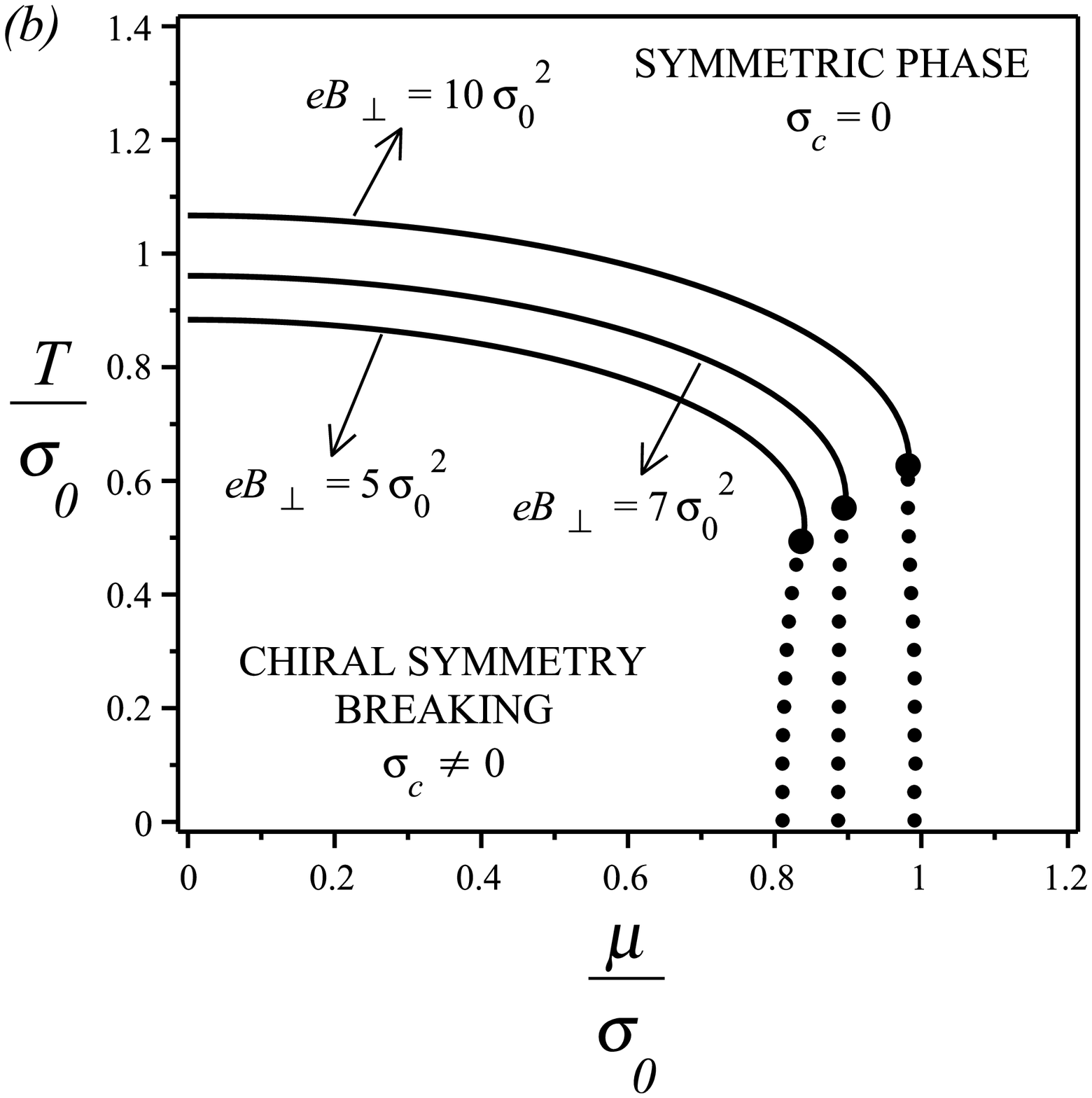,width=7.5cm}}
\caption{The phase diagram in the $(\mu,T)$ plane for $\lambda_R<0$.
  (a) The plot on the left shows the critical curves in the case of
  $B_\perp=0$ and for different values of $\delta \mu$.  (b) The plot on
  the right shows the critical curves for the case of $\delta \mu=0$
(in the absence of a Zeeman energy term) 
and for different values of $B_\perp$.  The solid
  lines are second-order critical lines and the dotted lines are
  first-order critical lines. The tricritical points are marked by the
  large dots in the plot on the right. All quantities are expressed in
  units of $\sigma_0 = \pi/|\lambda_R|$.}
  \label{fig-separateB}
\end{figure}

Let us initially show the effect on the phase structure of the system
when only one of the components of the magnetic field is applied to
the system, i.e., for the case of $\delta \mu\neq 0$, $B_\perp=0$ 
(i.e., when only the parallel component of the magnetic
field is present) and for the case when $\delta \mu=0$, but $B_\perp \neq 0$ 
corresponding to the case of neglecting the Zeeman energy term,
or, equivalently, when $B_\parallel= 0$ and the contribution
of $B_\perp $ through the Zeeman energy term is small enough such that it can
be neglected (which is the case in most of the relevant physical situations,
as discussed in Sec.~\ref{sec2} and exemplified in the Conclusions section).  
In {}Fig.~\ref{fig-separateB} we show each of these cases in terms of the 
phase diagram in the $(\mu,T)$ plane.  In
the two plots shown in the figure,  the external region to the curves
is a chiral-symmetry-restored phase  region, $\bar{\sigma}_c=0$, while
the internal region to the curves is a chiral-symmetry-broken phase
region, $\bar{\sigma}_c \neq 0$. {}For simplicity, we are not showing
the metastable regions that would appear around the first-order
transition lines in the plot on the right. These metastable regions
will be discussed and shown below. {}Figure \ref{fig-separateB}(a) is
for the case where the perpendicular component of the magnetic field
is absent, $B_\perp=0$. In this case, the critical lines for different 
values for the asymmetry $\delta \mu$ ($B_\parallel \neq 0$) 
are found to be all of
second order.  The only first-order transition point happens at $T=0$
and $\mu_c=\sigma_0$.  {}For a critical value of the asymmetry
$\delta\mu_c = \sigma_0$, we find that $\bar{\sigma}_c=0$ for any
value of the temperature.  Note that this critical value for the
asymmetry has the same value as for $\mu_c$, found in the absence of
the Zeeman energy term~\cite{rose}. This is a consequence of the
symmetry between the Zeeman energy term ($\delta \mu)$ and the
chemical potential.  {}Figure \ref{fig-separateB}(b) is for the case
where the in-plane component of the  magnetic field is absent,
$B_\parallel=0$, but $B_\perp \neq 0$ (and its contribution to
$\delta \mu$ is neglected).  The structure obtained in this case is
the one previously found in the earlier
papers~\cite{Klimenko,Gusynin}, where, besides the second-order
transition line, now a first-order transition line also appears, with
a tricritical point between them. The effect due to magnetic catalysis
is clear, showing that the larger $B_\perp$ is, the larger the
symmetry-broken region will be.  It is then clear from the results shown in
{}Fig.~\ref{fig-separateB} that the Zeeman energy term tends to
act to restore the chiral symmetry, while the perpendicular magnetic
field alone tends to enlarge the chiral-symmetry-broken  region.

\begin{figure}[htb]
\centerline{\psfig{file=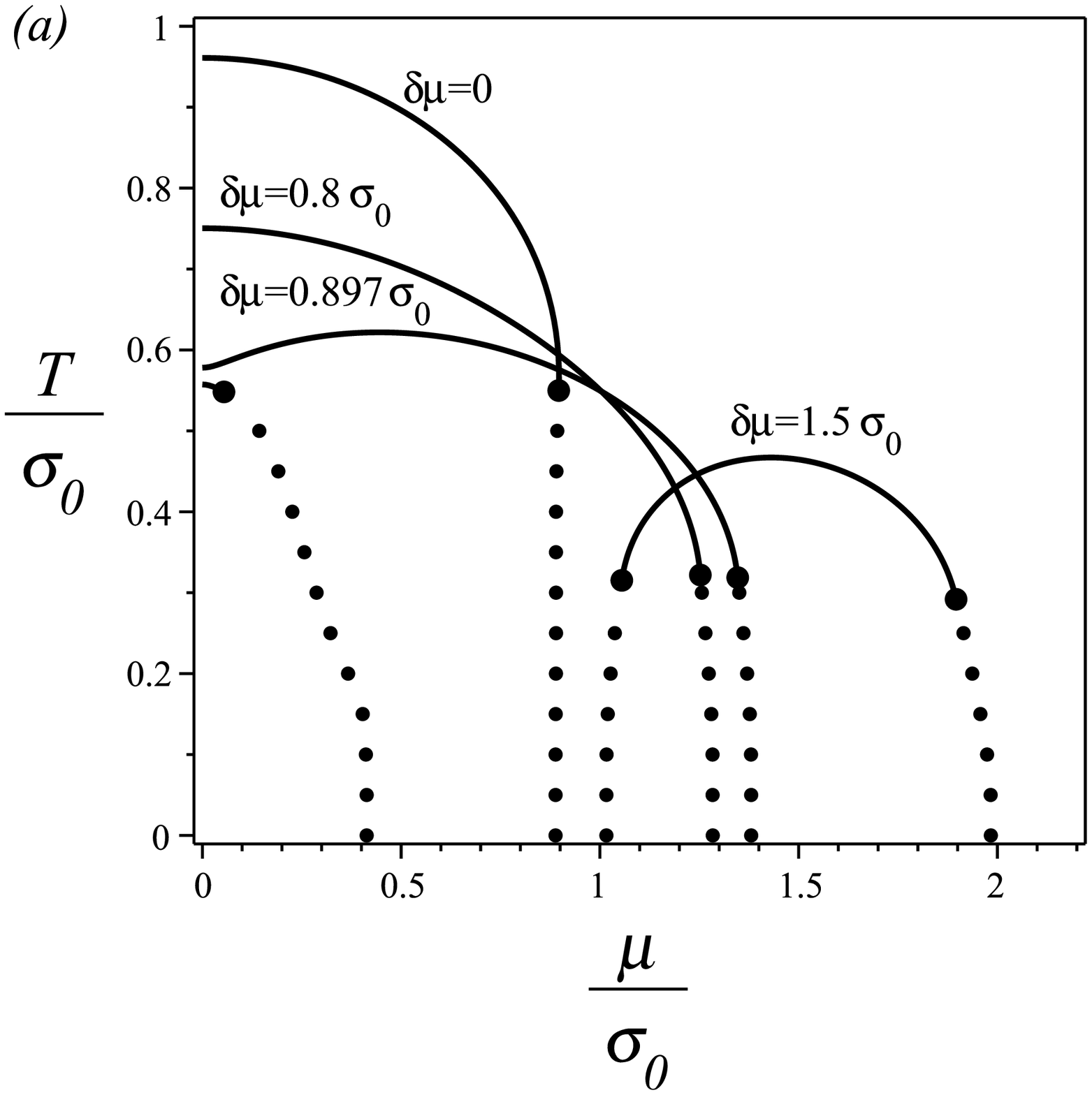,width=7.5cm}
\hspace{0.5cm} \psfig{file=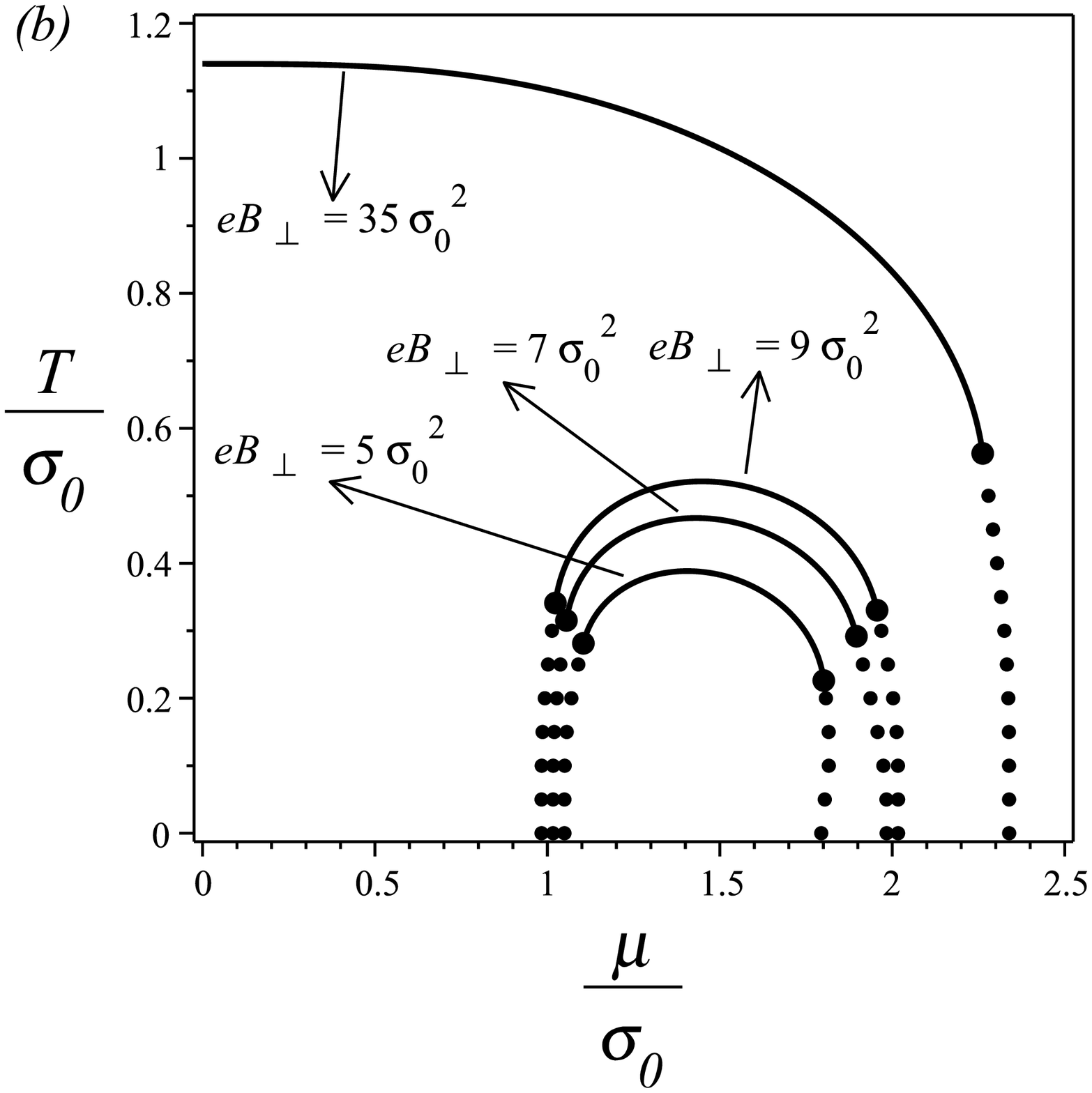,width=7.5cm}}
\vspace*{8pt}
\caption{The phase diagram in the $(\mu,T)$-plane for $\lambda_R<0$.
  (a) The plot on the left shows the critical curves in the case of fixed
  $|eB_\perp|=7 \sigma_0^2$ and for different values of $\delta \mu$.
  (b) The plot on the right shows the critical curves for the case of
  fixed $\delta \mu=1.5 \sigma_0$ and for different values of
  $|eB_\perp|$.  The solid lines are second-order critical lines and
  the dotted lines are first-order critical lines. The tricritical
  points are marked by the large dots in the plots.
  \label{fig2B}}
\end{figure}

\begin{figure}[htb]
\centerline{\psfig{file=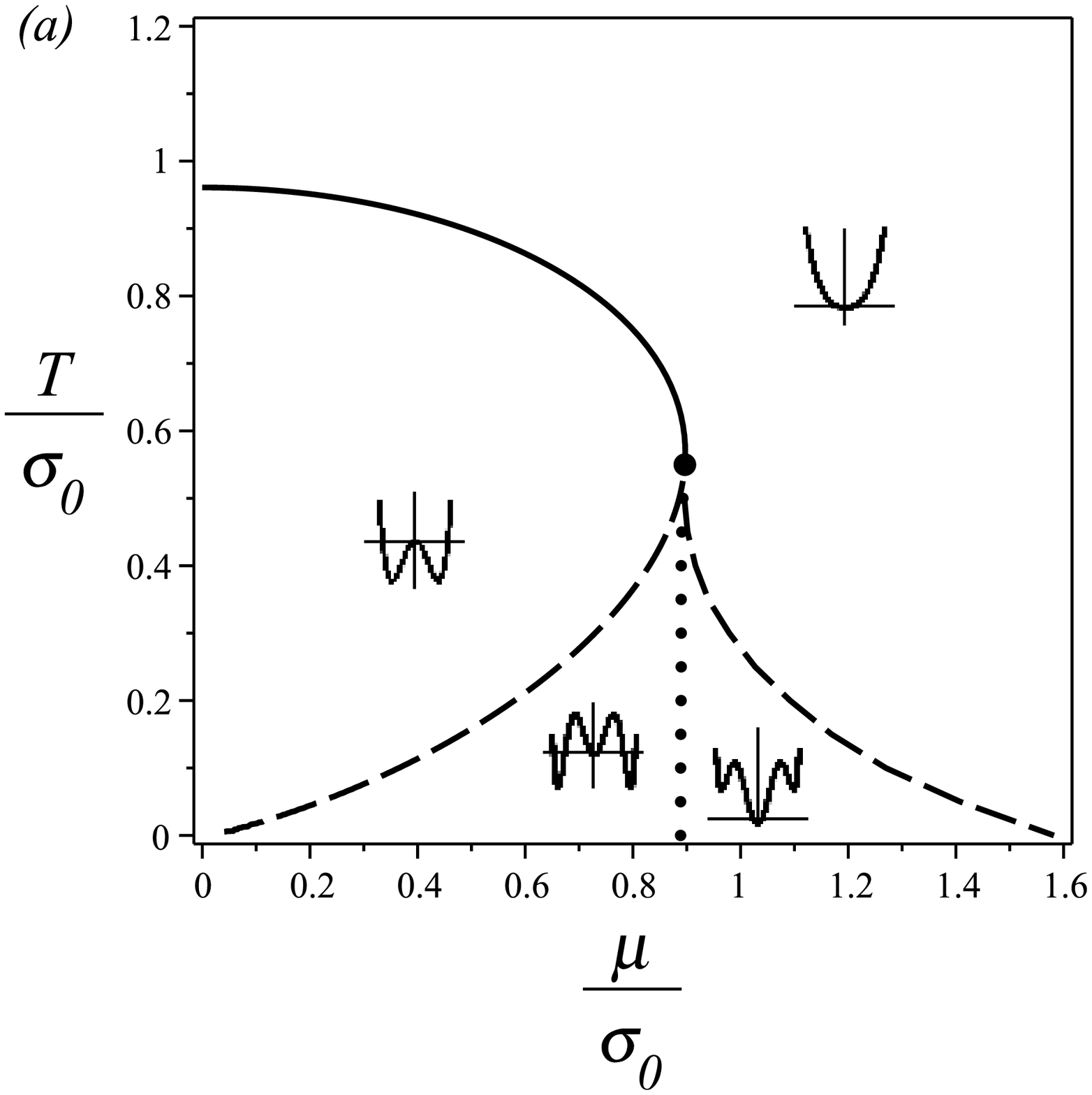,width=7cm}
\hspace{0.5cm} \psfig{file=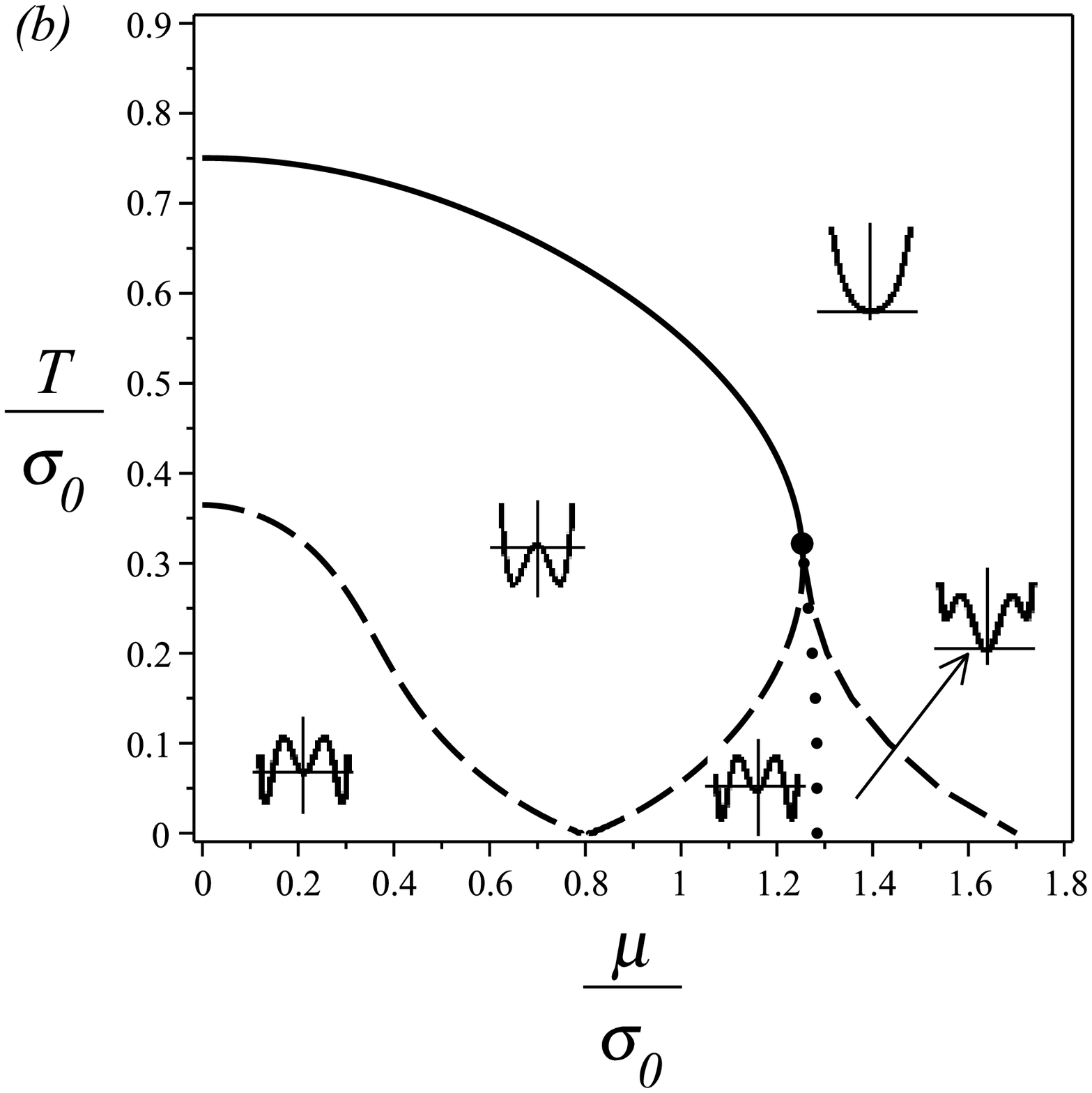,width=7cm}}
\vspace*{8pt} \centerline{\psfig{file=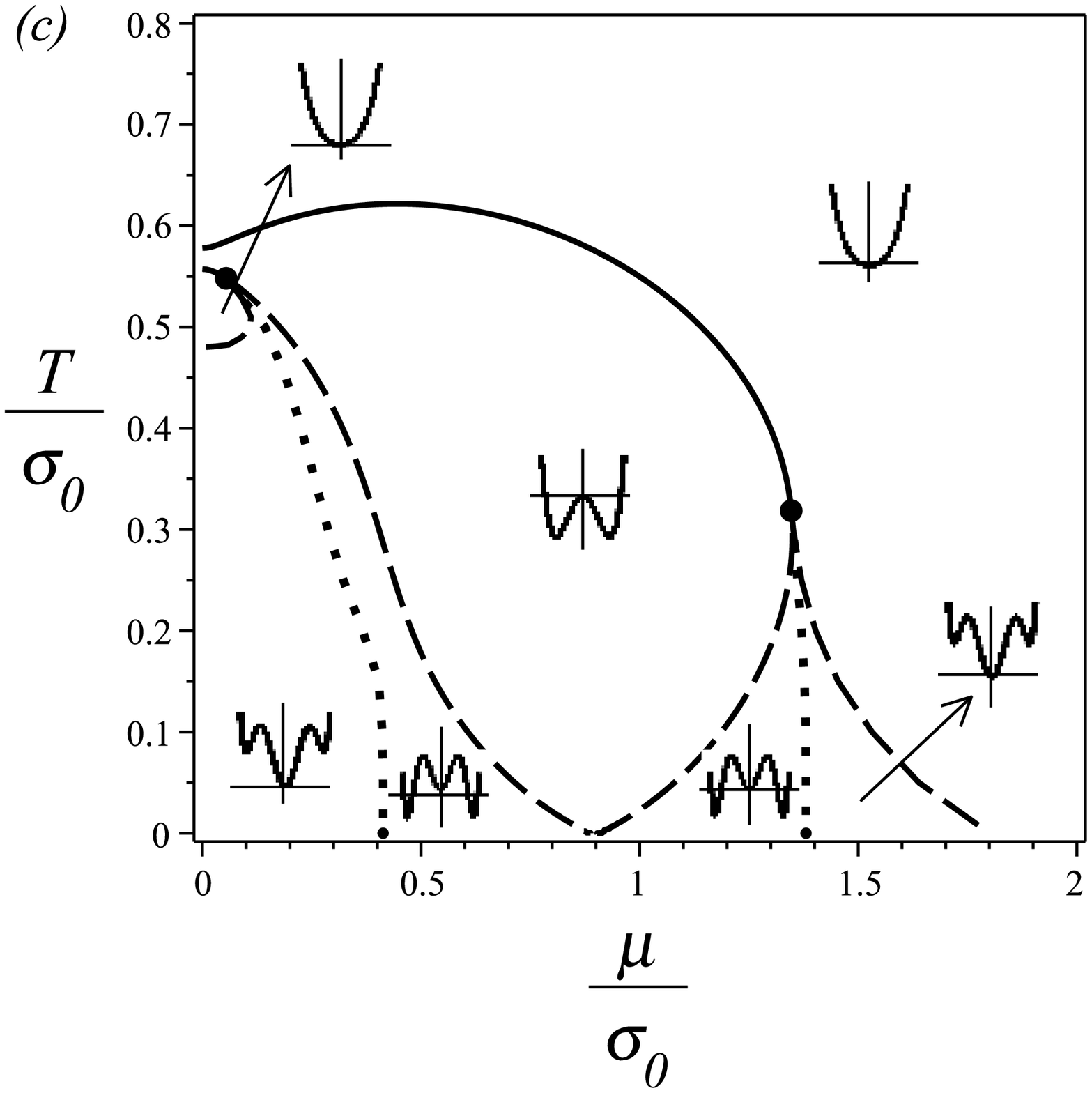,width=7cm}
\hspace{0.5cm} \psfig{file=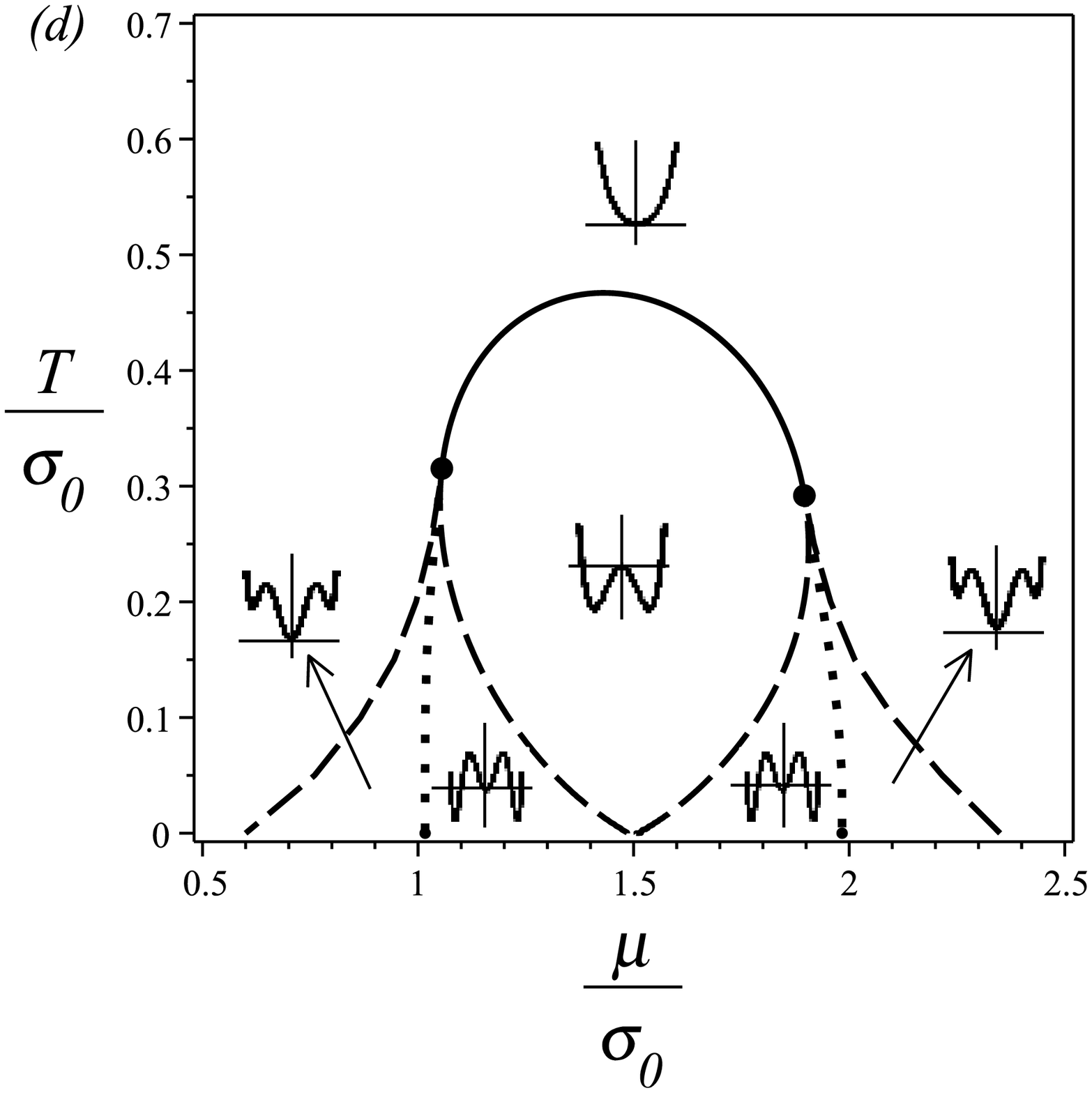,width=7cm}}
\caption{The phase diagram (including the metastable lines) in the
  $(\mu,T)$ plane.  The perpendicular magnetic field is fixed in
  $|eB_\perp|=7 \sigma_0^2$.  The panel (a) on the top left shows the
  critical curves in the case of $\delta \mu=0$.  The panel (b) on the
  top right shows the critical curves for the case of $\delta \mu=0.8
  \sigma_0$. The panel (c) on the  bottom left shows the critical
  curves for the case of $\delta \mu=0.897 \sigma_0$. The panel (d) on
  the bottom right shows the critical  curves for the case of $\delta
  \mu=1.5 \sigma_0$.  The solid lines are second-order critical lines,
  the dotted lines are first-order critical lines and the dashed lines
  are the  metastable critical lines. The tricritical points are
  marked by the large dots in the plots. Also shown is the schematic
shape of the effective potential in each region of the phase diagram.
  \label{fig-meta}}
\end{figure}

In {}Fig.~\ref{fig2B} we show the effects of increasing either 
the asymmetry $\delta \mu$ or the perpendicular component of the
magnetic field.  In {}Fig.~\ref{fig2B}(a) the perpendicular magnetic field  is
kept constant at the value of $|e B_\perp|=7 \sigma_0^2$ and 
the asymmetry $\delta \mu$
is varied.  In {}Fig.~\ref{fig2B}(b) we show the opposite case, where
we keep fixed the asymmetry $\delta \mu$ at the value 
$\delta \mu = 1.5 \sigma_0$, while the perpendicular
component of the magnetic field is varied.  A noticeable feature that
we now observe in each of these cases, not seen in the previous figure, 
is the possibility of the emergence of multiple  critical points and reentrant
phases.  In the case of fixed $B_\perp$, {}Fig.~\ref{fig2B}(a), after
some critical value for the asymmetry, which for the parameters taken
for this plot is $\delta \mu \simeq 0.895 \sigma_0$, a second
tricritical point emerges with an associated first-order transition
line.  In the plot shown in {}Fig.~\ref{fig2B}(b), we have purposively
considered a  value of asymmetry of $\delta \mu = 1.5
\sigma_0$. Recall that for an asymmetry above the critical value of
$\delta\mu_c = \sigma_0$ and $B_\perp=0$, the chiral symmetry
would always be restored~\cite{PRBinplane}.  However, as we start
increasing  $B_\perp$, a symmetry-broken region starts to emerge
after some large enough value for the perpendicular magnetic field.
This region of chiral-symmetry-broken phase that appears also has the
form of the multicritical curves shown in {}Fig.~\ref{fig2B}(a).
{}For a sufficiently large value of $B_\perp$, which for the
parameters used in the plot of {}Fig.~\ref{fig2B}(b) is $| e B_\perp|
\simeq 35 \sigma_0^2$, one of the additional tricritical points (and
its associated first-order transition line) disappears and the phase
diagram shows again a standard form with a second-order critical
line, a  first-order transition line and a tricritical point in
between them, a structure just like the ones 
shown in {}Fig.~\ref{fig-separateB}(b).
The combined effect of both the Zeeman energy term, generating the asymmetry,
and the perpendicular component of the magnetic field, producing the
Landau energy levels and associated magnetic catalysis, 
is able to lead to reentrant phase
transitions, as seen from the results in {}Fig.~\ref{fig2B}, whenever
more than one tricritical  point is present. This feature will be even
more apparent when we consider the zero-temperature limit and look at
the possibility of quantum phase transitions at fixed values of
$B_\perp$ and $\delta \mu$ and varying chemical potential, a
situation we will analyze in more detail below, in Sec. \ref{sec4}.

{}For completeness, in {}Fig.~\ref{fig-meta} we present the detailed
structure for some of  the phase diagrams shown in
{}Fig.~\ref{fig2B}, including now the metastable lines. The metastable
lines give the points for which a local minimum of the effective
potential first appears or disappears.  These lines are the ones shown
on the left and on the right of the first-order critical lines.  Again,
from the results shown in {}Figs.~\ref{fig-meta}(c) and
\ref{fig-meta}(d), we can see the presence of reentrant phase
transitions at some fixed temperature and varying chemical potential
(e.g., the analogue of varying the doping in condensed matter
systems). Reentrant phase transitions are always associated with more
than one critical point. Multiple critical points can appear with more
frequency as we decrease the temperature and, in particular, as we
enter the quantum regime ($T=0$). In the next section we will analyze
this situation more carefully.


\section{Quantum phase transition patterns}
\label{sec4}

In order to better understand the reentrant phases found in the
previous section, it is convenient to first look at those cases where
$T=0$, i.e., for the quantum regime. Taking the $T=0$ limit in
Eq.~(\ref{Veff}), we obtain 

\begin{eqnarray}
\frac{1}{N}V_{\rm eff,R}(\sigma_{c},T,\mu,B_\perp,\delta\mu)
&\stackrel{T\rightarrow 0} {\longrightarrow}&  \frac{\sigma_c^2}{2
  \lambda_R} -  \frac{\sqrt{2} |e B_\perp|^{3/2}}{\pi} \zeta\left(
-\frac{1}{2}, \frac{\sigma_c^2}{2 |e B_\perp|} +1 \right) -
\frac{|\sigma_{c}| |eB_{\perp}|}{2 \pi} \nonumber
\\ &-&\frac{|eB_{\perp}|}{4\pi}\left[\vphantom{\sum_{k=1}^{\infty}}
  \left(\mu_{\uparrow}-|\sigma_{c}|\right) \theta
  \left(\mu_{\uparrow}-|\sigma_{c}| \right) +
  \left(|\mu_{\downarrow}|-|\sigma_{c}|\right) \theta
  \left(|\mu_{\downarrow}|-|\sigma_{c}| \right) \right.  \nonumber
  \\ &+&\left.2 \sum_{k=1}^{k_{\rm max}^\uparrow}
  \left(\mu_{\uparrow}-\sqrt{2 k|e B_\perp| + \sigma_c^2} \right)
  \theta\left(\mu_{\uparrow}^2-\sigma_{c}^2 \right) \right. \nonumber
  \\ &+& \left. 2 \sum_{k=1}^{k_{\rm max}^\downarrow}
  \left(|\mu_{\downarrow}|-\sqrt{2k |e B_\perp| + \sigma_c^2} \right)
  \theta\left(\mu_{\downarrow}^2-\sigma_{c}^2 \right) \right] \, ,  
\label{VeffA}
\end{eqnarray}
where 

\begin{equation}
k_{\rm max}^{\uparrow, \downarrow} = {\rm
  Int}\left(\frac{\mu_{\uparrow,\downarrow}^2-\sigma_{c}^2}{2 |e
  B_\perp|}\right)\,,
\end{equation}
and ${\rm Int}(x)$ means the integer part of $x$.

\begin{figure}[htb]
\centerline{\psfig{file=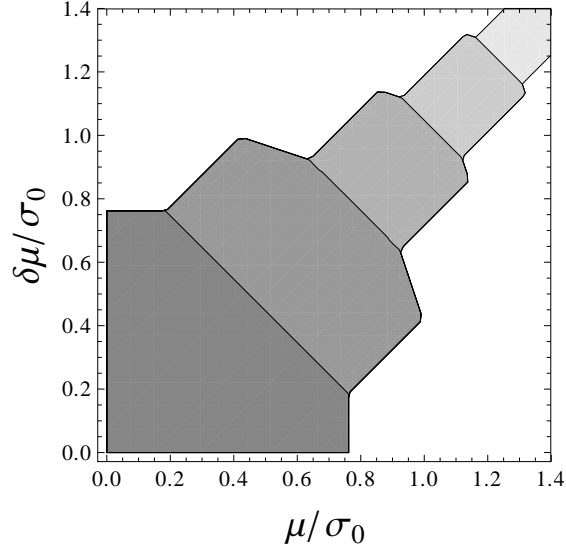,width=7.5cm}}
\caption{The phase diagram in the $(\mu,\delta \mu)$ plane and at
  $T=0$. The perpendicular magnetic field component was fixed at the
  value of $|eB_\perp|=\sigma_0^2$. The different shadows of gray
  indicate decreasing values for the vacuum expectation value
  $\bar{\sigma}_c$. Going from dark to  light shadows, we have
  $\bar{\sigma}_c/\sigma_0 \simeq 1.0617,\, 0.8144,\,
  0.3669,\,0.2883$. Outside the shadowed areas,
  $\bar{\sigma}_c=0$.  All solid lines indicate first-order phase
  transition lines.}
  \label{muXdeltamu}
\end{figure}

\begin{figure}[htb]
\centerline{\psfig{file=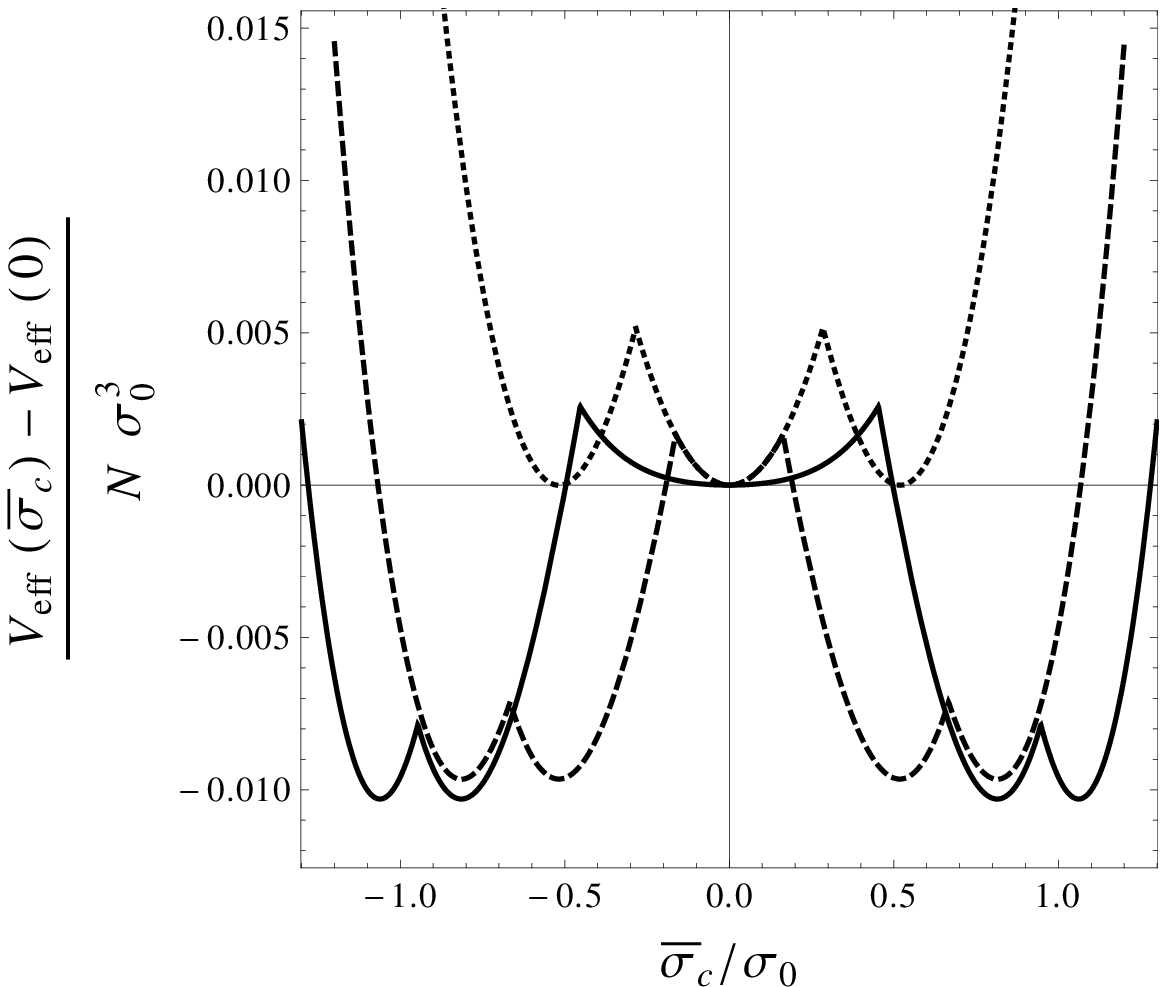,width=7.5cm}
\hspace{0.5cm}  \psfig{file=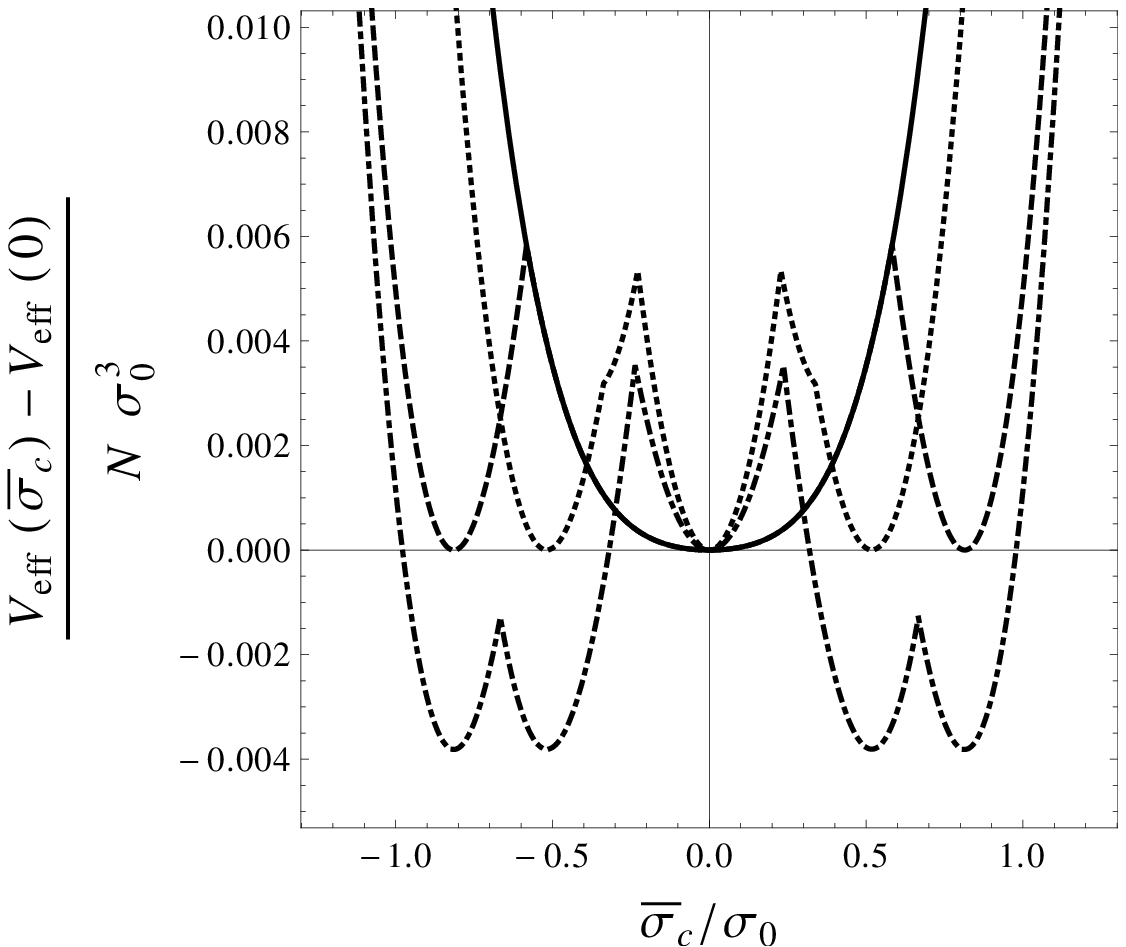,width=7.5cm}}
\caption{The effective potential for a fixed value of $|eB_{\perp}|=
  \sigma_{0}^{2}$ and for different values of chemical potential $\mu$
  and asymmetry $\delta \mu$.  The plot on the left is for $\delta
  \mu=0.7 \sigma_0$ and for  $\mu/\sigma_0\simeq 0.2467$ (solid
  curve), $\mu/\sigma_0\simeq 0.8633$ (dashed curve) and
  $\mu/\sigma_0\simeq 0.9845$ (dotted curve). {}For $\mu/\sigma_0 >
  0.9845$, there is only one global minimum at $\bar{\sigma}_c=0$ and
  chiral symmetry is restored.  The plot on the right is for $\delta
  \mu=0.9 \sigma_0$ and for  $\mu/\sigma_0 = 0$ (solid curve),
  $\mu/\sigma_0\simeq 0.3172$ (dashed curve), $\mu/\sigma_0\simeq
  0.6633$ (dash-dotted curve) and $\mu/\sigma_0\simeq 1.1282$ (dotted
  curve). {}For $\mu/\sigma_0 < 0.3172$ and for $\mu/\sigma_0 >
  1.1282$, there is only one global minimum at $\bar{\sigma}_c=0$ and
  chiral symmetry is restored.}
\label{figVeff}
\end{figure}

We can now look at the phase diagram in the $(\mu,\delta \mu)$ plane
at fixed $B_\perp$. An example is shown in {}Fig.~\ref{muXdeltamu}.
Note from {}Fig.~\ref{muXdeltamu} that for fixed values for the
asymmetry $\delta \mu$, we can find multiple
reentrant phases.  Examples of how these reentrant phases manifest
from the effective potential are shown in  {}Fig.~\ref{figVeff} for
two specific values of asymmetry, $\delta \mu = 0.7 \sigma_0$ and
$\delta \mu =0.9 \sigma_0$.

{}From the forms of the effective potential shown in
{}Fig.~\ref{figVeff}, we see that two forms of (first-order) reentrant
phase transitions are possible at some value for the asymmetry and when
the chemical potential is increased.  We can have a reentrant phase
transition starting at a chiral broken state, $\bar{\sigma}_c =
\sigma_1$, going to one or more intermediate chiral broken states with
$\bar{\sigma}_c = \sigma_2$, $\sigma_1>\sigma_2$, and then to a chiral
symmetric  state, $\bar{\sigma}_c=0$. This is the case shown, e.g., in
the first panel in {}Fig.~\ref{figVeff}. The other possibility is,
starting from a chiral symmetric state, $\bar{\sigma}_c=0$, going to one
or more intermediate chiral broken states, with $\bar{\sigma}_c \neq
0$, and then back to the chiral symmetric state. This is, e.g.,
represented in the second panel in {}Fig.~\ref{figVeff}.


\subsection{Quantum phase transitions in the fully polarized system}

\begin{figure}[htb]
\centerline{\psfig{file=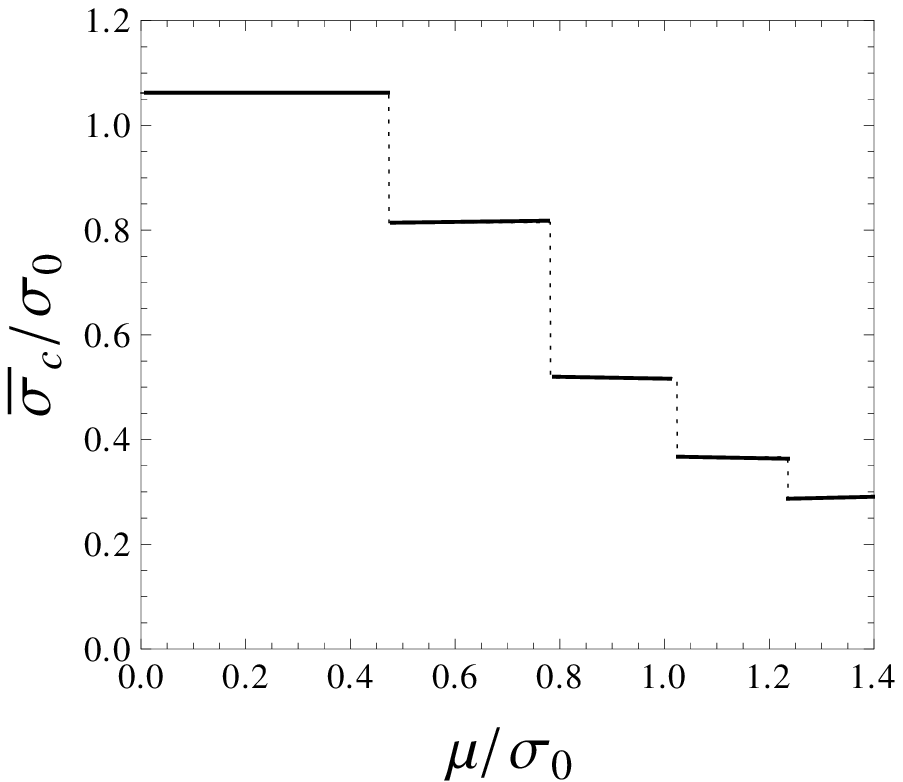,width=7.3cm}}
\caption{The values assumed for the vacuum expectation value
  $\bar{\sigma}_c$ along the line $\delta \mu = \mu$ in
  {}Fig.~\ref{muXdeltamu}.}
  \label{sigmamudeltamu}
\end{figure}

It is interestingly to notice from {}Fig.~\ref{muXdeltamu}  that the
reentrant phases are maximal (occur more frequently) when
$\mu=\delta\mu$. This corresponds  to the full polarization regime,
$\mu_\uparrow=2\mu$ and $\mu_\downarrow=0$. In
{}Fig.~\ref{sigmamudeltamu} we  show the values assumed for the vacuum
expectation value $\bar{\sigma}_c$ along this full polarization
regime. By increasing the chemical potential, the value of
$\bar{\sigma}_c$ can decrease discontinuously through many
intermediate first-order phase transitions. This type of pattern is
found to be exclusive of the full polarization regime. 

\begin{figure}[htb]
\centerline{\psfig{file=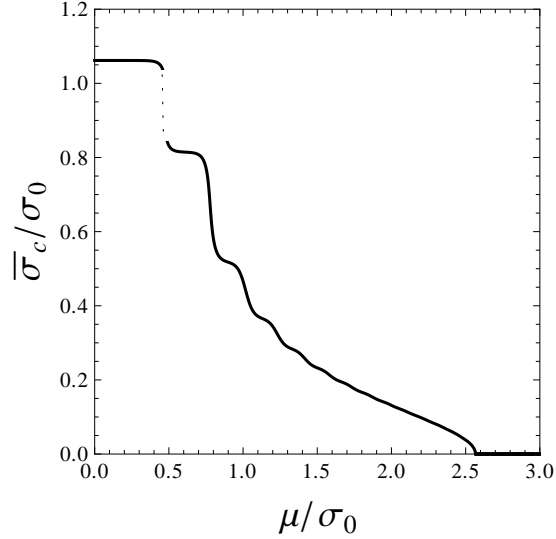,width=7.3cm}}
\caption{The values assumed for the vacuum expectation value
  $\bar{\sigma}_c$ along the line $\delta \mu = \mu$, when $|e
  B_\perp|=\sigma_0^2$ and $T= 0.06 \sigma_0$.}
  \label{sigmaTmudeltamu}
\end{figure}

Though this pattern of multiple phase transitions can happen at any
nonzero value of $B_\perp$ in principle, this structure is not stable
when thermal fluctuations are accounted for. Upon leaving the quantum regime
$T=0$ and by increasing the
temperature, the intermediate transitions going from a value of
$\bar{\sigma}_c =\sigma_1$ to a  value $\sigma_2$ (where $\sigma_1 >
\sigma_2$) quickly disappear.  This can be easily understood from the
results shown for the effective potential in
{}Fig.~\ref{figVeff}. These intermediate first-order transitions are
in general characterized by a small potential barrier. By increasing
the temperature, these intermediate minima in the potential are
quickly smoothed out. The smaller $B_\perp$ is, the smaller the
temperature needed to smooth out the intermediate transitions will be. {}For
example, in {}Fig.~\ref{sigmaTmudeltamu}, we have the case of $|e
B_\perp| = \sigma_0^2$ and for a temperature of $T=0.06 \sigma_0$,
which are taken as representative values.  All intermediate
transitions are smoothed out and $\bar{\sigma}_c$ becomes mostly  a
continuously varying function, with only the strongest
first-order transition remaining, which the first one  shown in
{}Fig.~\ref{sigmamudeltamu}, which also remains in
{}Fig.~\ref{sigmaTmudeltamu}.  By further increasing the temperature,
the transition becomes just a second-order one, with $\bar{\sigma}_c$
continuously varying before vanishing completely at a given critical
value of chemical potential. This corresponds to crossing one of the
lines of second-order phase transition.

{}From the expression for the effective potential in the $T=0$ limit,
Eq.~(\ref{VeffA}), we can find approximate analytical expressions for
the vacuum expectation value $\bar{\sigma}_c$ in the large magnetic
field regime, with $|e B_\perp| \gg \mu^2$ and $|e B_\perp| \gg
\bar{\sigma}_c^2$. {}For instance, in the $\mu=\delta \mu=0$ case, we
find, valid for either positive or negative values of $\lambda_R$,

\begin{equation}
\bar{\sigma}_c (T=0,\delta \mu=0,\mu=0) \simeq  \frac{\lambda_R |e
  B_\perp|} {2 \pi - \sqrt{2 | e B_\perp|} \zeta(1/2) \lambda_R}\;,
\label{sigmaT0mu0deltamu0}
\end{equation}
where $\zeta(1/2) \simeq -1.46035$, while for the case when there is
only the perpendicular component of the magnetic field, with
$B_\parallel=0$ and setting $\delta \mu=0$, we find

\begin{eqnarray}
\bar{\sigma}_c(T=0, \delta\mu=0,\mu)\simeq\begin{cases} 
\frac{\lambda_R |e B_\perp|} {2 \pi - \sqrt{2 | e B_\perp|} \zeta(1/2)
  \lambda_R}\ , &  \mu < \tilde{\mu}_{\rm c}\,,\\ 0 \ , & \mu \geq
\tilde{\mu}_{\rm c}\,,
\end{cases}
\end{eqnarray}
where

\begin{equation}
\tilde{\mu}_{\rm c} \simeq  \frac{\lambda_R |e B_\perp|} {4 \pi -
  2\sqrt{2 | e B_\perp|} \zeta(1/2) \lambda_R}\;.
\end{equation}
These results can be compared with the one in the fully polarized
case,  $\delta \mu = \mu\neq 0$. Taking $|e B_\perp| \gg \mu^2$ and
$\mu \gg \bar{\sigma}_c$, we find

\begin{equation}
\bar{\sigma}_c (T=0,\delta \mu=\mu) \simeq  \frac{\lambda_R |e
  B_\perp|} {4 \pi - 2\sqrt{2 | e B_\perp|} \zeta(1/2) \lambda_R}\;,
\label{sigmaT0mu=deltamu}
\end{equation}
which is half of the value obtained for $\bar{\sigma}_c$ when
$\mu=\delta\mu=0$,  Eq.~(\ref{sigmaT0mu0deltamu0}).

 
\subsection{Hall conductivity}
\label{hallsection}

It is not only convenient but also useful to look at possible physical
quantities that can be measured in the laboratory and give
confirmation of the existence of the reentrant and multiple phase
transitions we have found above. In most realistic experiments, the
measured quantities of interest are related, for example, to
susceptibilities, like the magnetic susceptibility and also quantities
related to electric transport.  In the presence of a tilted magnetic
field, we can have each field contributing differently to these
quantities. {}For example, it is expected that the enhancement
of the Zeeman splitting caused by the in-plane component of the
magnetic field will have important
effects in the experiments with graphene and also with  other planar
systems.  Based on experimental studies of an in-plane magnetic field
in  graphene, the Zeeman splitting has been shown to be important in
both the spin transport and conductance fluctuation
properties~\cite{folk}. It has also been shown that the Zeeman
splitting leads to the spectrum of the effective single-particle
Hamiltonian exactly as required by the observed pattern of
quantization of Hall  conductivity~\cite{Kim}. Since the Hall
conductivity is a typical quantity of  importance for analysis when
studying these systems, we will here concentrate on the possible
effects the reentrant and multiple phase transitions we have found can
have on it. 

The Hall conductivity, $\sigma_H$, can be defined as~\cite{Kittel}

\begin{equation}
\sigma_H = \frac{e^2}{| e B_\perp|} n\,,
\label{sigmaH}
\end{equation}
where $n$ is the number density,

\begin{eqnarray}
n= -\frac{\partial}{\partial\mu}V_{\rm
  eff,R}(\sigma_{c},T,\mu,B_\perp,\delta\mu)
\Bigr|_{\sigma_{c}=\bar{\sigma}_{c}} \, .
\end{eqnarray}

{}From the zero-temperature limit for the effective potential,
Eq.~(\ref{VeffA}), we find

\begin{eqnarray}
\sigma_{H}(\bar{\sigma}_{c},T=0,\mu,B_{\perp},\delta \mu)&=&
\frac{Ne^2}{4\pi} \left[1+2
  \ \mathrm{Int}\left(\frac{\mu_{\uparrow}^{2}-
    \bar{\sigma}_{c}^2}{2|eB_{\perp}|}\right)\right]
\theta(\mu_{\uparrow}-|\bar{\sigma}_{c}|)
\nonumber \\ &+& \frac{Ne^2}{4\pi} {\rm sign}(\mu_{\downarrow})
\left[1+2 \ \mathrm{Int}\left(\frac{\mu_{\downarrow}^{2}-
    \bar{\sigma}_{c}^2}{2 |e B_{\perp}|}\right)\right]
\theta(|\mu_{\downarrow}|-|\bar{\sigma}_{c}|) \ , 
\label{Hallfull}
\end{eqnarray} 
where ${\rm sign}(x)$ is the sign function. 

\begin{figure}[H]
\centerline{\psfig{file=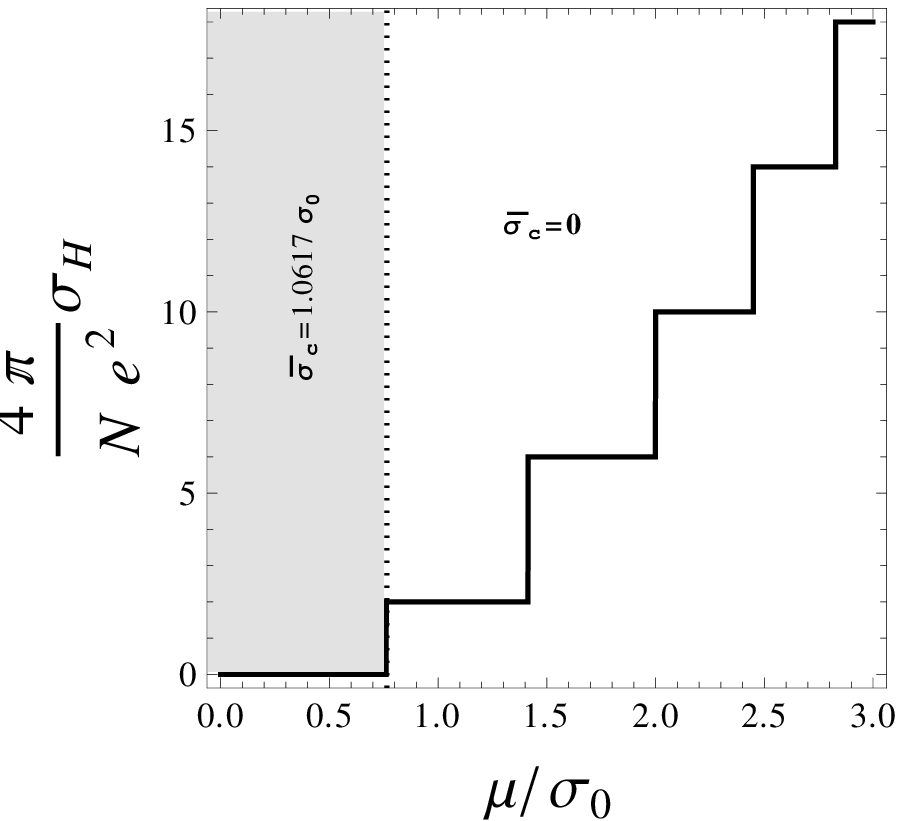,width=6.5cm}
\hspace{0.5cm} \psfig{file=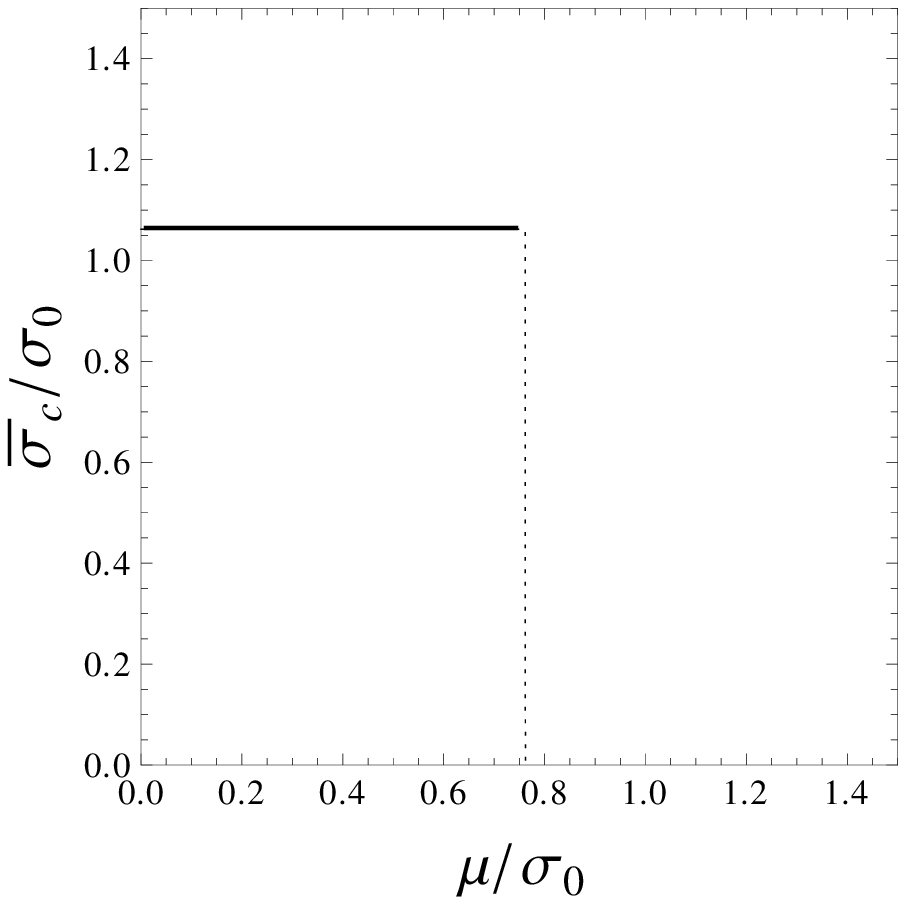,width=6cm}}
\vspace*{5pt}
\centerline{\psfig{file=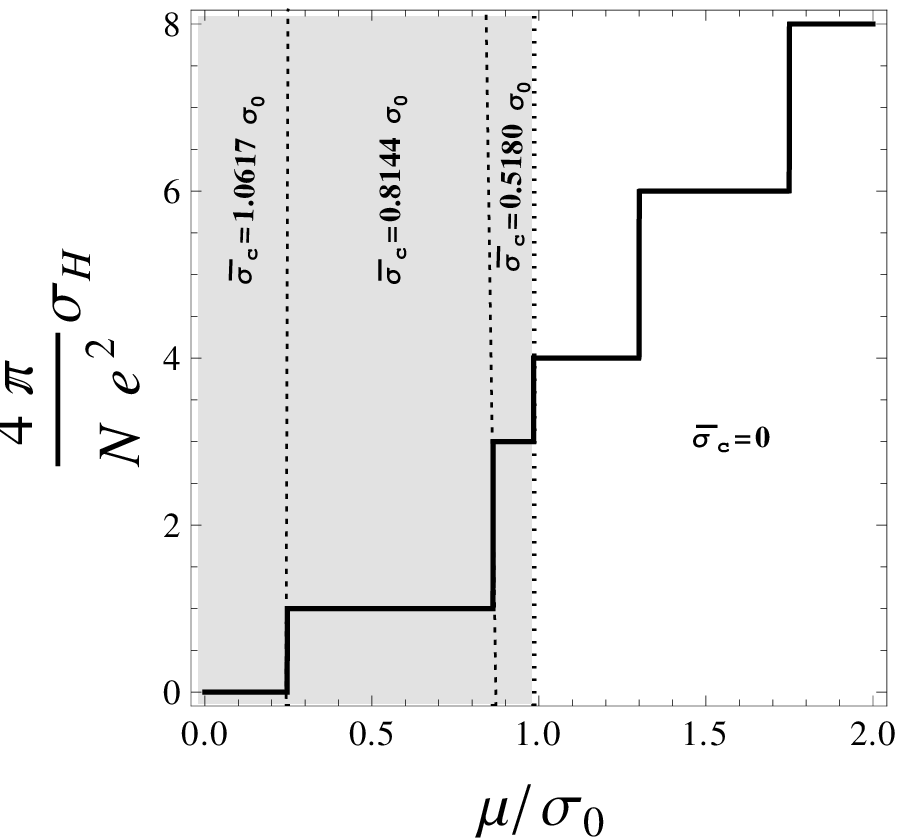,width=6.5cm}
\hspace{0.5cm} \psfig{file=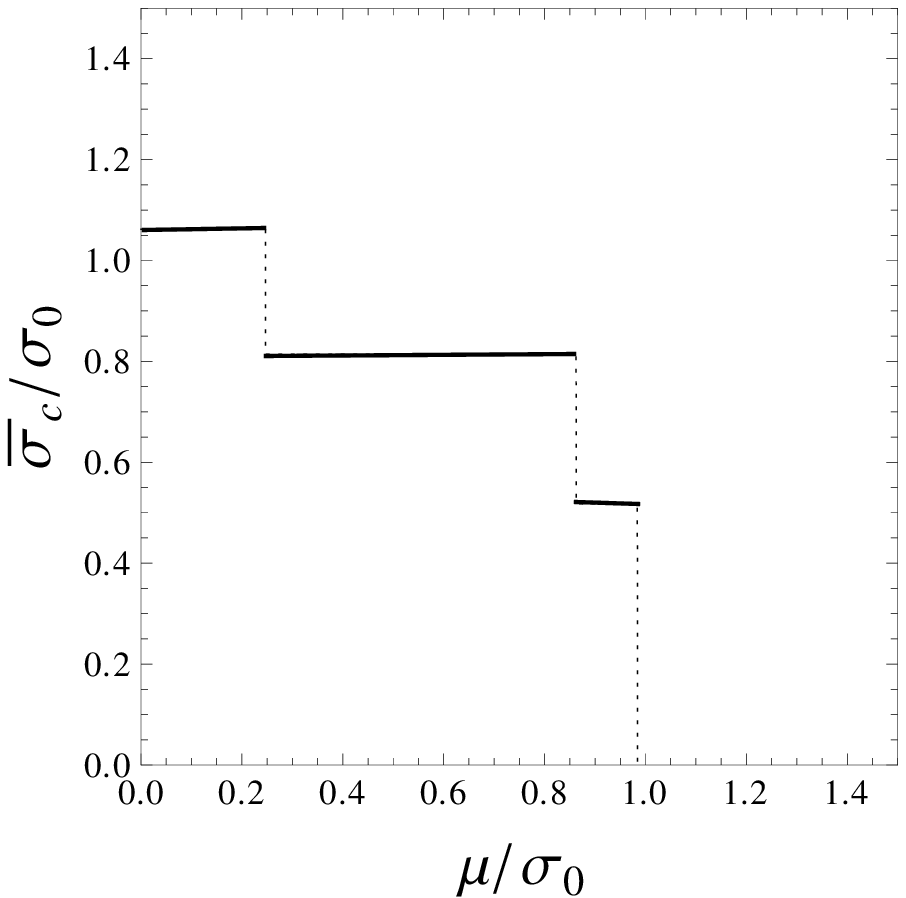,width=6cm}}
\vspace*{5pt}
\centerline{\psfig{file=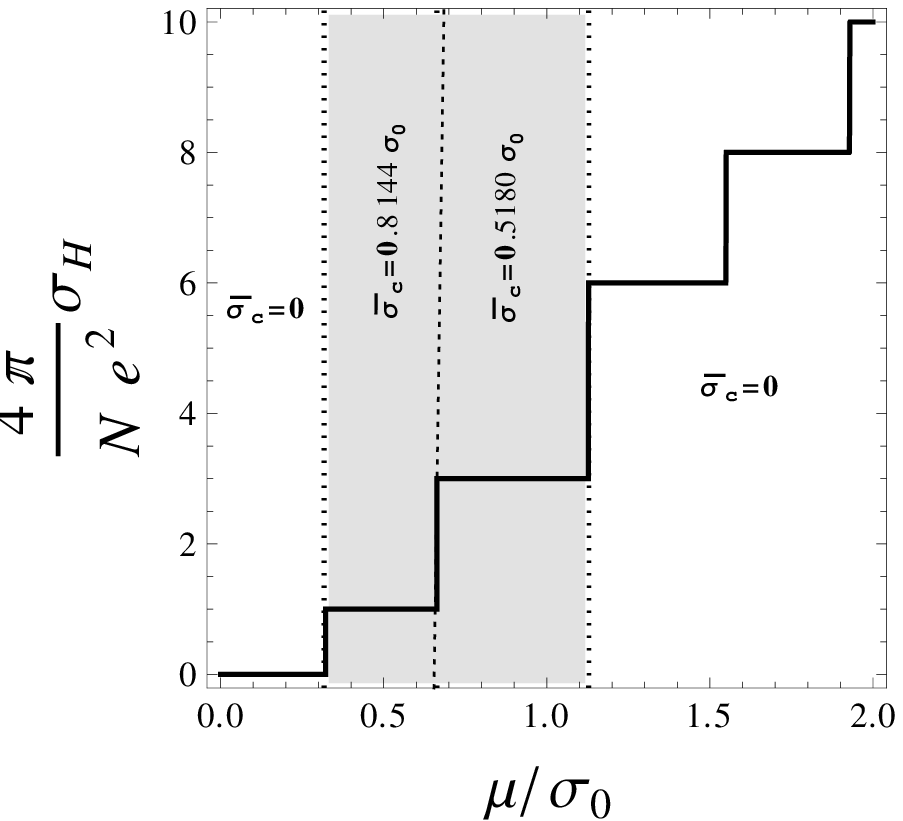,width=6.5cm}
\hspace{0.5cm} \psfig{file=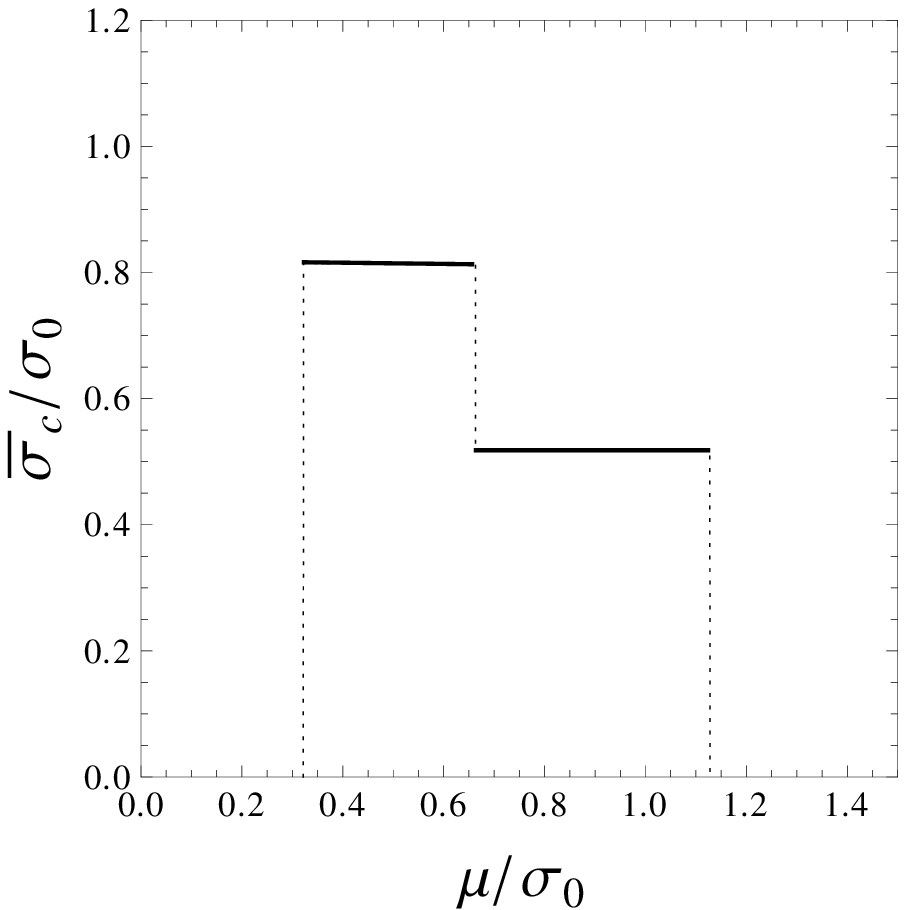,width=6cm}}
\caption{The Hall conductivity, at zero temperature, as a  function of
  chemical potential (plots on the left). The vertical thin dotted
  lines separate the regions for which $\bar{\sigma}_{c}$ changes
  value (discontinuously).  The plots on the right show the
  corresponding variation of  $\bar{\sigma}_{c}$ with the chemical
  potential.  {}From top to bottom we have the cases of $\delta \mu=0$,   
$\delta\mu=0.7\sigma_{0}$ and
  $\delta\mu=0.9\sigma_{0}$.  The perpendicular component of the
  magnetic  field is held fixed at the value $|eB_{\perp}|=
  \sigma_{0}^{2}$.  }
\label{figHall}
\end{figure}

In {}Fig.~\ref{figHall} we show some representative cases for the Hall
conductivity at $T=0$, as a function of the chemical potential and for
different values for the  asymmetry $\delta \mu$.  We have again kept
fixed, for convenience, the perpendicular component of the magnetic
field in the value $|eB_{\perp}|=\sigma_{0}^{2}$ as a representative
example. The vertical dotted lines in the plots on the left indicate
the positions of the phase transitions, with the shaded areas
indicating the chiral-symmetry-breaking regions and the corresponding
values assumed for $\bar{\sigma}_c$.  The pattern of chiral phase
transition in each case is shown in the plots on the right.  Note that
each time there is a transition, the Hall conductivity can jump by a
different  factor than in the absence of chiral symmetry breaking.  As
the chemical potential is increased from zero, the largest differences
appear  at the first transition and at the last transition, when the
system enters in the chiral symmetric  region. This behavior is quite
different from that seen when computing the Hall conductivity in
quantum electrodynamics (see, e.g., Ref.~\cite{Martinez}) where there
are no gaps and transitions,  i.e., where $\bar{\sigma}_{c}=0$. The
presence of a gap $\bar{\sigma}_{c}$ and its capacity to change as the
chemical potential is changed, as seen in the previous  subsection, is
responsible for the differences in the Hall conductivity.  This is
determined  by the relations between $\mu_{\uparrow,\downarrow}$ and
$\bar{\sigma}_{c}$, for a given value of $B_\perp$, as can be inferred
from Eq.~(\ref{Hallfull}).  Recalling that
$\mu_{\uparrow}=\mu+\delta\mu$ and $\mu_{\downarrow}=\mu-\delta\mu$,
we find for the Hall conductivity the following behaviors  as we vary
the values of $\mu$ and $\delta \mu$:

\begin{itemize}

  \item When $\mu_{\downarrow}\geq0$:

   \begin{eqnarray}
    \frac{4\pi}{Ne^2}\sigma_{H}(\bar{\sigma}_{c},T=0,\mu,B_{\perp},\delta
    \mu) &=&\left[1+2 \ \mathrm{Int}\left(\frac{\mu_{\uparrow}^{2}-
        \bar{\sigma}_{c}^2}{2|eB_{\perp}|}\right)\right]
    \theta(\mu_{\uparrow}-|\bar{\sigma}_{c}|) \nonumber
    \\ &+&\left[1+2 \ \mathrm{Int}\left(\frac{\mu_{\downarrow}^{2}-
        \bar{\sigma}_{c}^2}{2|eB_{\perp}|}\right)\right]
    \theta(|\mu_{\downarrow}|-|\bar{\sigma}_{c}|) \ .
\end{eqnarray}

\begin{itemize}

\item If $|\mu_{\downarrow}|>|\bar{\sigma}_{c}|$, we always have that
  $\mu_{\uparrow}>|\bar{\sigma}_{c}|$ and

\begin{eqnarray}
 \frac{4\pi}{Ne^2}\sigma_{H}(\bar{\sigma}_{c},T=0,\mu,B_{\perp},\delta
 \mu) &=& 2 \left[1+\mathrm{Int}\left(\frac{\mu_{\uparrow}^{2}-
     \bar{\sigma}_{c}^2}{2|eB_{\perp}|}\right)+
   \mathrm{Int}\left(\frac{\mu_{\downarrow}^{2}-
     \bar{\sigma}_{c}^2}{2|eB_{\perp}|}\right)\right].
\label{caso1}
\end{eqnarray}

\item If $\mu_{\uparrow}>|\bar{\sigma}_{c}|$ and
  $|\mu_{\downarrow}|<|\bar{\sigma}_{c}|$, then

\begin{eqnarray}
 \frac{4\pi}{Ne^2}\sigma_{H}(\bar{\sigma}_{c},T=0,\mu,B_{\perp},\delta
 \mu) &=& 1+\mathrm{Int}\left(\frac{\mu_{\uparrow}^{2}-
   \bar{\sigma}_{c}^2}{2|eB_{\perp}|}\right). 
\label{caso2}
\end{eqnarray}

\item If $\mu_{\uparrow}<|\bar{\sigma}_{c}|$, we always have that
  $|\mu_{\downarrow}|<|\bar{\sigma}_{c}|$, then

\begin{eqnarray}
 \frac{4\pi}{Ne^2}\sigma_{H}(\bar{\sigma}_{c},T=0,\mu,B_{\perp},\delta
 \mu) &=& 0. \label{caso3}
\end{eqnarray}

\end{itemize}

\end{itemize}

 \begin{itemize}

  \item When $\mu_{\downarrow}<0$:

   \begin{eqnarray}
    \frac{4\pi}{Ne^2}\sigma_{H}(\bar{\sigma}_{c},T=0,\mu,B_{\perp},\delta
    \mu) &=&\left[1+2 \ \mathrm{Int}\left(\frac{\mu_{\uparrow}^{2}-
        \bar{\sigma}_{c}^2}{2|eB_{\perp}|}\right)\right]
    \theta(\mu_{\uparrow}-|\bar{\sigma}_{c}|) \nonumber
    \\ &-&\left[1+2 \ \mathrm{Int}\left(\frac{\mu_{\downarrow}^{2}-
        \bar{\sigma}_{c}^2}{2|eB_{\perp}|}\right)\right]
    \theta(|\mu_{\downarrow}|-|\bar{\sigma}_{c}|) \ .
\end{eqnarray}

\begin{itemize}

 \item If $|\mu_{\downarrow}|>|\bar{\sigma}_{c}|$, we always  have
   that $\mu_{\uparrow}>|\bar{\sigma}_{c}|$, then:

\begin{eqnarray}
 \frac{4\pi}{Ne^2}\sigma_{H}(\bar{\sigma}_{c},T=0,\mu,B_{\perp},\delta
 \mu) &=& 2 \left[\mathrm{Int}\left(\frac{\mu_{\uparrow}^{2}-
     \bar{\sigma}_{c}^2}{2|eB_{\perp}|}\right)-
   \mathrm{Int}\left(\frac{\mu_{\downarrow}^{2}-
     \bar{\sigma}_{c}^2}{2|eB_{\perp}|}\right)\right]. 
\label{caso4}
\end{eqnarray}

  \item If $\mu_{\uparrow}>|\bar{\sigma}_{c}|$ and
    $|\mu_{\downarrow}|<|\bar{\sigma}_{c}|$, we have that:

\begin{eqnarray}
 \frac{4\pi}{Ne^2}\sigma_{H}(\bar{\sigma}_{c},T=0,\mu,B_{\perp},\delta
 \mu) &=& 1 + \mathrm{Int}\left(\frac{\mu_{\uparrow}^{2}-
   \bar{\sigma}_{c}^2}{2|eB_{\perp}|}\right). 
\label{caso5}
\end{eqnarray}

  \item If $\mu_{\uparrow}<|\bar{\sigma}_{c}|$, we always have that
    $|\mu_{\downarrow}|<|\bar{\sigma}_{c}|$ and

\begin{eqnarray}
 \frac{4\pi}{Ne^2}\sigma_{H}(\bar{\sigma}_{c},T=0,\mu,B_{\perp},\delta
 \mu)&=&0. 
\label{caso6}
\end{eqnarray}

\end{itemize}

\end{itemize}

Equations (\ref{caso5}) and (\ref{caso6}) are identical to
Eqs.~(\ref{caso2}) and (\ref{caso3}), respectively (the sign of
$\mu_{\downarrow}$ does not interfere in these cases).  Thus, we have
four distinct possibilities for the Hall conductivity,  given 
by  Eqs. (\ref{caso1}), (\ref{caso2}), (\ref{caso4}))
and (\ref{caso6}). 

\begin{figure}[htb]
\centerline{\psfig{file=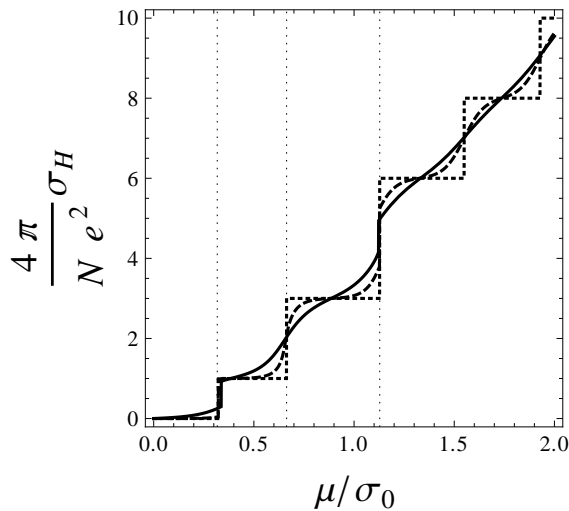,width=7.5cm}}
\caption{The Hall conductivity for $|eB_\perp|=\sigma_0^2$, $\delta
  \mu=0.9 \sigma_0$  and for $T=0$ (dotted line), $T=0.05 \sigma_0$
  (dashed line) and $T=0.1 \sigma_0$ (solid line).}
  \label{HallfiniteT}
\end{figure}

We also note that by accounting for the effect of thermal
fluctuations, we will have two effects on the Hall
conductivity. {}First, as seen in the  previous subsection, it will
tend to smooth out the intermediate transitions and only the strongest
first-order phase transitions tend to remain at a given
temperature. Second, including the effect of temperature, the Hall
conductivity itself is smoothed out. This is shown in
{}Fig.~\ref{HallfiniteT}, for the example of asymmetry $\delta \mu=0.9
\sigma_0$ that was considered in {}Fig.~\ref{figHall}.  Note that for
the values of temperature considered, the intermediate transition
disappears and  only the first and last transitions remain, the one
entering the chiral-symmetry-broken region and the one leaving
it, respectively. 

\section{Conclusions}
\label{sec5}

In this work we have studied how an external tilted magnetic field,
with  nonvanishing components parallel and perpendicular to the
system's plane,   can affect the phase diagram for planar fermions in
a GN type of model with discrete chiral symmetry.  Though in this work
we have concentrated on the transition patterns found for the chiral
fermionic condensate, we can easily extend our work  to include also a
superconducting gap~\cite{Klimenko-recent}. We will investigate this
in detail elsewhere.

As far the chiral symmetry is concerned, the parallel component of the
magnetic field leads to a Zeeman effect in  these planar systems that
tends to weaken the chiral symmetry. Thus,  the chiral symmetry phase
transition, which in condensed matter systems can be associated with
an insulating-to-metal type of transition,  may happen at a smaller
critical temperature in the presence of a Zeeman field. This behavior,
due to an increasing magnetic Zeeman field, is exactly the opposite of
that observed when a perpendicular magnetic field is applied to the
system (see e.g. Refs.~\cite{Klimenko,Gusynin}). In a perpendicular magnetic
field the chiral symmetry breaking becomes stronger as a consequence
of the effect of the magnetic field, a behavior associated with the
magnetic catalysis in the system. (Note that magnetic catalysis is not
an effect exclusive to fermionic systems but can also occur in
bosonic systems as well~\cite{phiB}, enlarging the symmetry-broken
region in those systems).   Thus,  in the presence of a perpendicular
magnetic field the transition happens at a higher  critical
temperature.  We have shown that this competing effect of each
component of the external magnetic field applied to the system can
produce a rich phase diagram, with the production of multiple critical
points  and reentrant phase transitions. 

Note that there are many systems that can display multiple phase
transitions, most particularly in low-energy condensed matter systems,
where they can be found more easily. At the quantum field theory
level,  reentrant phases have been shown to be possible in systems
with coupled scalar fields in the nonrelativistic
limit~\cite{Ramos:2006er}.  Multiple phase transitions can in
principle also be found in relativistic systems. This can be the case,
for example,  in systems with two or more coupled scalar
fields (see, for example, Ref.~\cite{Pinto:1999pg} and references therein).  
Multiple critical points in effective
models for quantum chromodynamics have also been shown to be possible
in Ref.~\cite{Ferroni:2010ct}. However, to our knowledge, no reentrant
phase transitions of the type we have found here have been shown
before to appear in fermionic relativistic systems. 

As a way to determine the presence of multiple critical points and
reentrant phase transitions, we have computed the Hall
conductivity. We have seen that whenever a first-order phase
transition happens, the Hall conductivity jumps by a different amount
at the critical point. This is a consequence of the different value
assumed (discontinuously) for the chiral vacuum expectation value
$\bar{\sigma}_c$, as analyzed in Sec.~\ref{hallsection}.  Since
the Hall conductivity $\sigma_H$ is an important measurable quantity
for diverse condensed matter systems, like for semiconductor
materials, superconductor films  and graphene, our results indicate a
way that the presence of a gap in the system and quantum phase
transitions may reflect on the measurements of $\sigma_H$.  

It is tempting to try to directly apply our results to the physics of
graphene, for instance. In these systems we can have both components
of the magnetic  field acting concomitantly. However, in most studies
(see, e.g., Ref.~\cite{hallsplitting})  the Zeeman energy term (which for
us here is given by the asymmetry term $\delta \mu$) is much smaller
than the energy splitting between Landau levels. 
This is easily understood if we use typical values
found in realistic planar condensed matter systems, including,
for example, graphene. Recall that the Zeeman energy is
$\delta \mu = g \mu_B B/2$, where $B$ is the total magnitude of the magnetic field.
Using $g = 2$, $\mu_B = e/(2 m_e)$, where $m_e =
0.511$~MeV is the electron rest mass, $e \simeq
1/\sqrt{137}$ and the conversion factor 1 Tesla $\simeq 692.4 \,{\rm
  eV}^2$, we find, by expressing $B$ in terms of the perpendicular component of the
field and the tilt angle $\phi$ (see {}Fig.~\ref{BBfig1}), that

\begin{equation}
\delta \mu \simeq 0.058 \, \frac{B_\perp[\rm T]}{\sin \phi}~{\rm meV},
\label{dmu}
\end{equation} 
where $B[\rm T]$ is the absolute value of the magnetic
field in units of teslas.  {}For the Landau energy level splitting between the
zeroth and first levels, we find that

\begin{equation}
E_{LL} = \sqrt{2 v_F^2 | e
  B_\perp|} \simeq 36.3 \sqrt{B_\perp[\rm T]}~{\rm meV},
\label{ELL}
\end{equation} 
where we have reestablished the Fermi velocity in the above expression and used that $v_F
\sim c/300$ (e.g., like for graphene). 
We then find 
for the ratio $\delta \mu/E_{LL} \simeq 0.0016 \sqrt{B_\perp[\rm T]}/\sin \phi$.
The Zeeman energy term $\delta \mu$ is then comparable to $E_{LL}$ only for extremely
large values of the magnetic field (e.g., for $B_\perp \gtrsim  3.9\times 10^5$ Teslas,
when taking for instance that $B_\parallel=0$), or for a highly tilted magnetic field,
such that $\sin \phi \ll 1$. 
{}For typical laboratory
magnetic fields of ${\cal O}(10)$ Teslas or for a not-too-much tilted
magnetic field (such that $B_\perp \gtrsim B_\parallel$),  we find that
$\delta \mu \ll E_{LL}$. So, indeed, under these circumstances, the
Zeeman energy term becomes negligible and also justifies neglecting it
when $B_\parallel =0$. 
 
{}For the type of phenomena we have found in this paper, we can
already find reentrant and multiple phase transitions for  asymmetries
as small as $\delta \mu \sim 0.1 \sigma_0$ and at a temperature
$T \lesssim 0.01 \sigma_0$. Taking a perpendicular
component for the magnetic field of $v_F^2|e B_\perp| \sim 10
\sigma_0^2$  (where we have reestablished again the Fermi velocity in
the expression for the Landau energy levels, 
as appropriate for fermionic condensed matter systems)
and for a typical gap energy $\sigma_0$ of around $10\,$ meV
\cite{gapvalues}, we find $T\lesssim 1$ K and a required field 
with $B_\perp \sim 1.5$ Teslas and, upon using Eq.~(\ref{dmu}),
we obtain $\sin \phi \sim 0.087$, which then gives $B_\parallel \sim 17.2$ Teslas.  
This corresponds to a
highly tilted magnetic field in the direction of the system's plane,
which is the expected physical situation where the effects of the Zeeman energy 
term start to become important and cannot be neglected, according to the
estimates shown above (for these values of parameters, the Zeeman energy 
term $\delta \mu$ is about $2.3\%$ of the value of $E_{LL}$).
These estimates for the required values of magnetic fields seem
reasonable and low enough to be achieved in the laboratory and
this does not seem such a highly  special situation to be produced in
practice.  Besides, with the increasing precision being reached in the
most recent experiments, it may be possible to realize such an
experiment with superconductor films under these tuned conditions and
probe the type of transitions we have found in this paper, for example
through the measurements of the Hall conductivity, as we have
suggested.

In conclusion, we have seen that the opposite effects caused by
magnetic fields, when applied parallel and perpendicular to the
system's plane, can lead to a rich pattern of phase transitions.  The
results we have found here can have immediate applications in the
context of  the physics of planar condensed matter systems, including
graphene, planar films of conducting polyacetylene and
high-temperature superconductor films.  In all these systems, the
viability of their application as modern semiconductor devices is
directly related to the ability of producing and controlling a gap.
Our results indicate that this control can be achieved through  an
appropriate doping of the system and under parallel and perpendicular
magnetic fields, that are applied simultaneously and independently,
when properly tuned.  If these conditions can be achieved, then our
results can open interesting  possibilities for the uses of these type
of materials in practical applications  as novel electronic devices. 


\acknowledgments

The authors thank K. G. Klimenko for useful comments and suggestions.
R.O.R. is partially supported by research grants from Conselho
Nacional de Desenvolvimento Cient\'{\i}fico e Tecnol\'ogico (CNPq) and
Funda\c{c}\~ao Carlos Chagas Filho de Amparo \`a Pesquisa do Estado do
Rio de Janeiro (FAPERJ). P.H.A.M. is supported by Coordena\c{c}\~ao de
Aperfei\c{c}oamento de Pessoal de N\'{\i}vel Superior (CAPES).


\end{document}